\documentclass[prr,a4paper,reprint,superscriptaddress,preprintnumbers]{revtex4-2}
\pdfoutput=1

\usepackage{mymacros}
\definecolor{change}{rgb}{0.0, 0.0, 0.0} 
\newcommand{\change}[1]{{\textcolor{change}{#1}}}
\usepackage{booktabs}
\usepackage{hyperref}

\begin{document}

\title{From spin chains to real-time thermal field theory using tensor networks}

\author{Mari~Carmen~Ba\~nuls} \email{mari.banuls@mpq.mpg.de}
\affiliation{Max Planck Institute of Quantum Optics, 
85748 Garching bei M{\"u}nchen, Germany}
\affiliation{Munich Center for Quantum Science and Technology (MCQST), 
80799 M{\"u}nchen, Germany}

\author{Michal P.\ Heller} \email{michal.p.heller@aei.mpg.de}
\affiliation{Max  Planck  Institute  for  Gravitational  Physics (Albert Einstein Institute), 
14476 Potsdam-Golm, Germany}
\affiliation{National Centre for Nuclear Research, 
02-093 Warsaw, Poland}

\author{Karl Jansen} \email{Karl.Jansen@desy.de}
\affiliation{NIC, DESY-Zeuthen, 
15738 Zeuthen, Germany}

\author{Johannes Knaute} \email{johannes.knaute@aei.mpg.de}
\affiliation{Max  Planck  Institute  for  Gravitational  Physics (Albert Einstein Institute), 
14476 Potsdam-Golm, Germany}
\affiliation{Department of Physics, Freie Universit\"{a}t Berlin, 
14195 Berlin, Germany}

\author{Viktor Svensson} \email{viktor.svensson@aei.mpg.de}
\affiliation{National Centre for Nuclear Research, 
02-093 Warsaw, Poland}
\affiliation{Max  Planck  Institute  for  Gravitational  Physics (Albert Einstein Institute), 
14476 Potsdam-Golm, Germany}


\begin{abstract}
One of the most interesting directions in theoretical high-energy and condensed matter physics is understanding dynamical properties of collective states of quantum field theories. The most elementary tool in this quest are retarded equilibrium correlators governing the linear response theory. In this article we examine tensor networks as a way of determining them in a fully ab initio way in a class of (1+1)-dimensional quantum field theories arising as infra-red descriptions of quantum Ising chains. We show that, complemented with signal analysis using the Prony method, tensor network calculations for intermediate times provide a powerful way to explore the structure of singularities of the correlator in the complex frequency plane and to make predictions about the thermal response to perturbations in a class of non-integrable interacting quantum field theories.
\end{abstract}

\maketitle

\section{Introduction and motivation}

Much of the progress in quantum field theory (QFT) to date has been driven by challenges posed by quantum chromodynamics (QCD). One motivation behind this work comes from ultrarelativistic heavy-ion collisions, which probe collective states of QCD in a non-equilibrium setting~\cite{Busza:2018rrf}. Important insights in this context have been gained by studying small perturbations of equilibrium in soluble models, such as holography or kinetic theory~\cite{Florkowski:2017olj}. Another motivating aspect of our work stems from thermalization and relaxation being important and timely areas of quantum-many body physics~\cite{Nandkishore:2014kca, eisert2015quantum, Gogolin:2016hwy, DAlessio:2016rwt, Wilming_2018, RevModPhys.91.021001}.

Linear response theory provides a natural point of departure to study real-time dynamics of QFTs and is the subject of the present article. In this framework, the response of an equilibrium state to a (small) perturbation triggered by a source $\cal J$ coupled to an operator $\cal O$ is captured by the retarded thermal two-point correlator
\begin{equation}
G_{R}^{\cal O}(t,x) = i \, \theta(t) \, \mathrm{tr} \left\{ \rho_{\beta} [{\cal O}(t,x), {\cal O}(0,0)] \right\},
\end{equation}
where $\rho_{\beta}$ is the thermal density matrix at inverse temperature~$\beta$ and $\theta(t)$ the Heaviside function. It is useful to express the change in the expectation value of~${\cal O}$ in Fourier space~\footnote{Note that we distinguish functions from their Fourier transforms only through arguments.}
\begin{equation}
\delta \langle {\cal O}(t,p) \rangle = \int d\omega \, e^{- i \, \omega \, t} \, G_{R}^{\cal O}(\omega, p) \, {\cal J}(-\omega, - p),
\end{equation}
since non-analytic features of $G_{R}$ summarize important features of the response. For instance, for holographic QFTs~\cite{Birmingham:2001pj, Son:2002sd, Kovtun:2005ev}, the function has simple poles which give rise to terms exponentially decaying with time. In the QCD-like context, these terms can be used to identify transport coefficients 
as well as the time needed for hydrodynamics to apply. In the condensed matter studies, the exponential decay of magnetization with time for small perturbations of equilibrium in the thermal Ising model~\cite{sachdev_2011} at criticality can be traced to a sequence of single pole singularities in the Fourier transform of certain correlator~\cite{PhysRevB.87.245107,DAlessio:2016rwt,Sabella-Garnier:2019tsi}. In addition to single poles there may be also branch cut singularities~\cite{Hartnoll:2005ju,Romatschke:2015gic,Kurkela:2017xis, Moore:2018mma}, which often give rise to power-law behaviour in time.

Beyond weak and strong coupling QFTs, very little is known about time dependence of collective states in QFTs. In this article we explore the use of tensor networks (TNs)~\cite{Cirac2009rg, Verstraete2008, Schollwoeck2011, Orus2014a, Silvi2019tns,Banuls2018lat, Banuls2019review, Banuls2019qtflag}, in particular matrix product operator (MPO) methods, to study thermal retarded correlators in complexified frequency space. We focus on a class of (1+1)-dimensional QFTs arising as the infra-red (IR) description of quantum Ising models~\cite{Zamolodchikov:1989fp,Delfino2005Jul,Fonseca:2006au,kjall2011bound,Zamolodchikov:2013ama}, which include non-integrable interacting theories hosting a spectrum of non-perturbative bound states. We cross-check our numerics against exact QFT predictions using the Ising model in a parameter region in which the IR is described by a free massive fermion QFT.
We then study the quantum Ising model in the presence of a longitudinal magnetic field. We cover both the case of interacting integrable QFT arising at vanishing transverse magnetic field, and a generic non-integrable case involving both the transverse and longitudinal magnetic fields \cite{Zamolodchikov:1989fp,Delfino2005Jul,Fonseca:2006au,Zamolodchikov:2013ama}. We focus on the \mbox{dimension-1} fermion bilinear operator and on a homogeneous perturbation ($p = 0$) and use the so-called Prony method~\cite{peter2014generalized} to extract the complex frequency plane structure of correlators from the numerically-obtained real-time signal.

\section{Ising model and IR QFTs}

The quantum Ising model is given by the Hamiltonian
\begin{equation}  \label{eq:Isingham}
H = - J \,\left ( \sum_{j=1}^{L-1} \sigma_z^j \sigma_z^{j+1} + h \sum_{j=1}^L  \sigma_x^j + g \sum_{j=1}^L \sigma_z^j \right) ,
\end{equation}
where $J$ sets an overall energy scale, $L$ denotes the total number of spins (sites), $\sigma_{x, \, z}^j$ are Pauli matrices at position $j$ and $g$ / $h$ stand for the longitudinal / transverse (magnetic) field. When $L \rightarrow \infty$, there exists a scaling limit~\footnote{See appendix~\ref{app:A} in the supplemental material.} such that the IR (long distances w.r.t.\ the lattice spacing $a \equiv 2/J$) description of the Hamiltonian~\eqref{eq:Isingham} is given by the following Majorana fermion QFT \cite{Rakovszky:2016ugs}:
\small
\begin{equation}
\label{eq:H_IsingQFT}
\hspace{-4 pt} H_\text{IR} = \int_{-\infty}^\infty dx \,  \big\{ \frac{i}{4\pi} \left( \psi\partial_x\psi - \bar\psi\partial_x\bar\psi \right) - \frac{i M_h}{2 \pi}\bar\psi\psi + {\cal C} M_{g}^{15/8} \, \sigma \big\} ,
\end{equation}
\normalsize
where ${\cal C} \approx 0.062$. The parameters $M_{h}$ and $M_{g}$ have dimensions of mass and are related to the ones in eq.~\eqref{eq:Isingham} via $M_{h} \equiv 2 J |1-h|$ and $M_{g} \equiv \, {\cal D} J \, |g|^{8/15}$ with ${\cal D} \approx 5.416$ \cite{Rakovszky:2016ugs, Hodsagi:2018sul}. The scaling limit, in which the QFT description appears, corresponds to taking $M_h/J \to 0$ while keeping the ratio $M_{h}/M_{g}$ fixed, see ref.\,\cite{Rakovszky:2016ugs} for a discussion of the QFT limit. In addition, when temperature $\beta^{-1}$ is included, to stay in the continuum limit we require $\beta \, J~\gg~1$. 

If $M_{h} = M_{g} = 0$, then the Hamiltonian describes a free Majorana fermion (Ising) CFT with central charge equal to~$\frac{1}{2}$. Operators $i \, \bar{\psi} \psi$ and $\sigma$ are scalar primary Hermitian operators in this CFT of dimensions $\Delta = 1$ and $\Delta = \frac{1}{8}$ respectively. On the lattice, as is apparent from the relation between Hamiltonians~\eqref{eq:Isingham} and~\eqref{eq:H_IsingQFT}, these operators are  proportional to $\sigma_{x}^{j}$ (with proportionality constant equal simply to $-a/\pi$) and $\sigma_{z}^j$ respectively. 

For the above class of relevant deformations of the Ising CFT, there are two kinds of deformations which give rise to integrable QFT. For $M_h\neq0$, $M_g=0$ one gets a free fermion QFT with the fermion mass being $M_h$. In this case the same continuum theory can be represented by the spin chain in both the ferro- ($h<1$) and para-magnetic ($h>1$) phase. For $M_h=0$, $M_g\neq0$ one obtains the interacting integrable E$_8$ QFT~\cite{Zamolodchikov:1989fp}, which has 8 massive and stable particles -- fermionic bound states. The mass of the lightest, $M_{1}$, is given precisely by $M_g$ and the ratios of masses of the heavier particles are shown in table~\ref{tab:e8_masses_ratio}. 

The general form of eq.~\eqref{eq:H_IsingQFT} with both $i\, \bar{\psi} \psi$ and $\sigma$ turned on gives rise to an interacting non-integrable QFT which contains stable and unstable bound states~\cite{Zamolodchikov:1989fp,Delfino2005Jul,Fonseca:2006au,Zamolodchikov:2013ama}. We study this regime at non-zero temperature numerically using MPO methods in combination with Prony analysis, which we review later.

\section{Retarded thermal correlators in solvable cases}

For the transverse field Ising model, i.e. $g = 0$ in eq.~\eqref{eq:Isingham}, and in the limit of an infinite chain $L \rightarrow \infty$ one can use the free fermion formulation to find a simple formula for the retarded thermal correlator of the transverse magnetization~$\sigma_{x}^{j}$. At $p = 0$ one gets for $t>0$
\small
\begin{equation} \label{eq:TFIMcorrelator}
G_{R}^{- \frac{\pi}{a} \, \sigma_{x}} = 2J \int_{-\pi}^\pi dk \, (2 n_k-1)\, \sin^{2}{\phi_k} \,\sin{(2 \, \varepsilon_k \, t)} ,
\end{equation}
\normalsize
where the phases $\phi_k$ satisfy $\tan{\phi_k} = \frac{\sin{k}}{h-\cos{k}}$, $\varepsilon_k = 2 J\sqrt{(1 + h^2-2h \cos{k})}$ and $n_k = (1 + e^{\beta \, \varepsilon_k})^{-1}$ is the Fermi-Dirac distribution. There is an intuitive understanding of eq.~\eqref{eq:TFIMcorrelator} in terms of two-particle exchanges: the integral in eq.~\eqref{eq:TFIMcorrelator} encapsulates $\sigma_{x}^{j}$ exciting a continuum of states, which are distinguished by the relative momentum of the two fermions from which they are built.

\begin{figure}[!t]
\centering
 \includegraphics[width=1\linewidth]{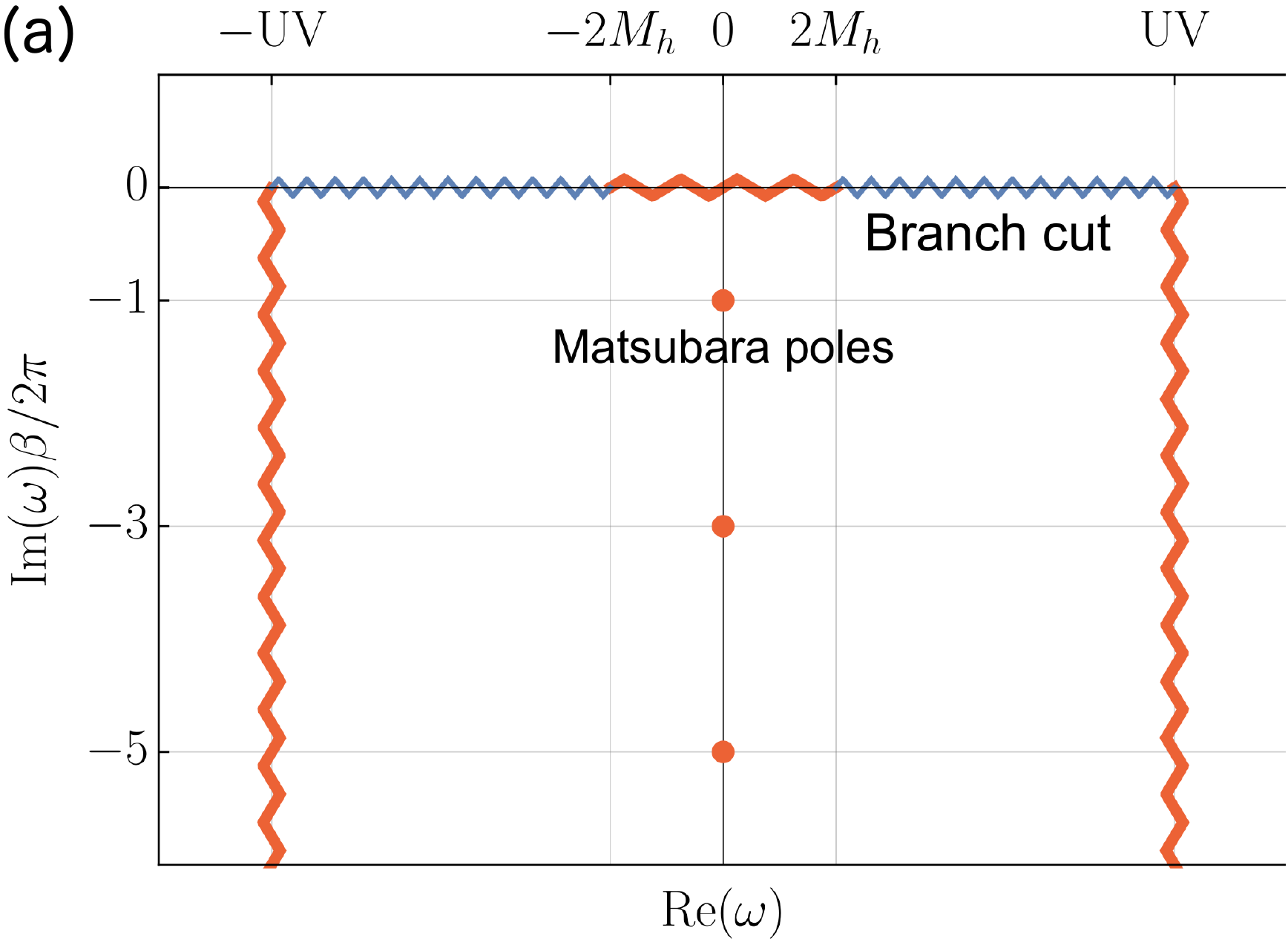}
 \includegraphics[width=0.995\linewidth,trim=-9mm 0 0 0]{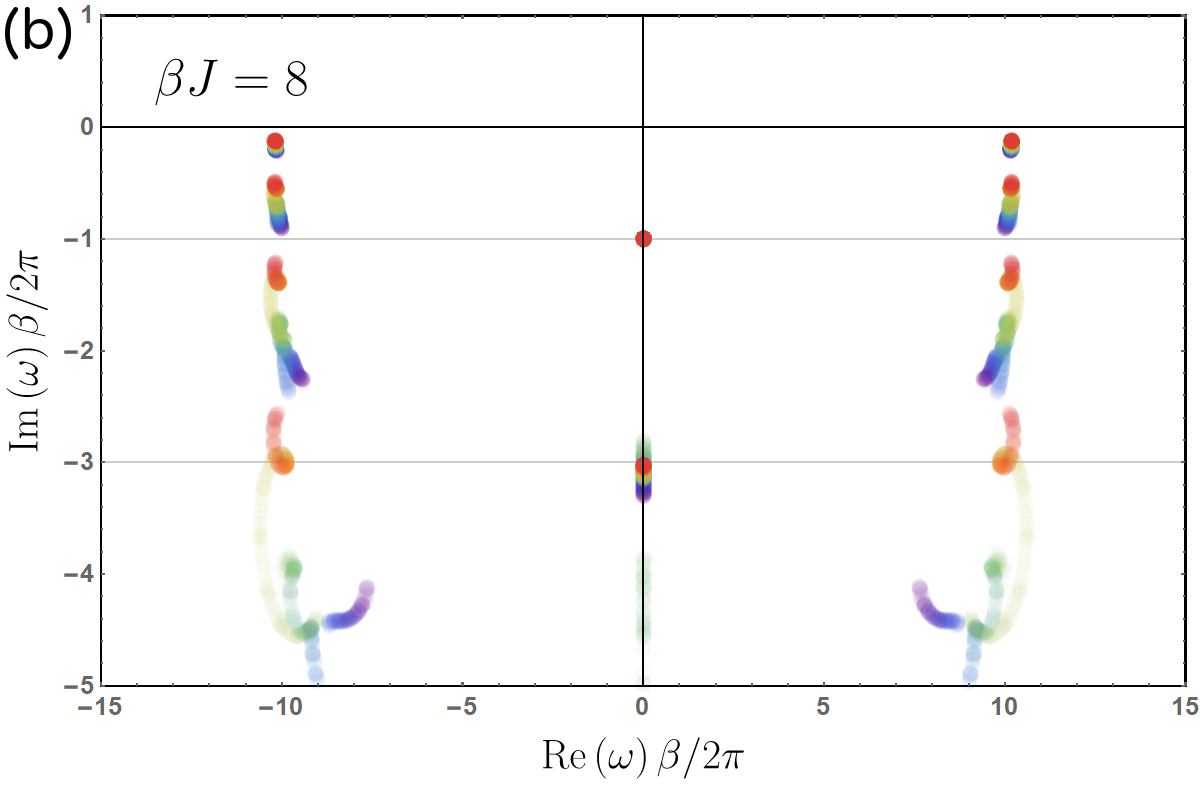}
 \includegraphics[width=0.8\linewidth]{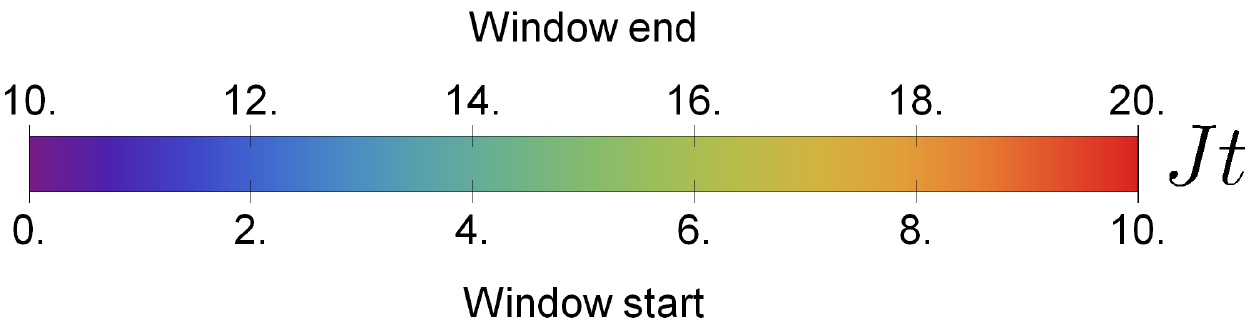} 
  \caption{The expected (a) and seen by the Prony method (b) analytic structure of eq.~\eqref{eq:TFIMcorrelator}. Branch cuts make it ambiguous and two choices are illustrated by blue and red. The latter  decouples the UV and the IR which makes it more natural for QFT and in this case single poles related to Matsubara frequencies arise. The Prony method applied to the case $M_h = 0$ prefers vertical branch cuts. Signal analysis is done for many different time windows, characterized by the color legend. The results for each window are concatenated into a single plot. The fuzzier a given pole is, the less robust the result. Branch cuts are represented by time-dependent poles, which leads to colorful bands. Prony identifies both the Matsubara poles as well as the UV branch cuts.}
 \label{fig:TFIM_correlator}
 \end{figure}

The analytic structure of eq.~\eqref{eq:TFIMcorrelator} in the complex frequency space is ambiguous, as noted in a similar context in ref.\,\cite{Kurkela:2017xis}. The natural representation is a branch cut stretching from the IR scale, set by twice the fermion mass~$M_{h}$, to the ultra-violet (UV) scale, set by the lattice spacing $a = 2/J$. 
\change{This is apparent from eq.~\eqref{eq:TFIMcorrelator}, which expresses the retarded correlator as a (continuous) sum over terms that oscillate in time with frequencies ranging from $2\epsilon_0=2 M_h$ to $2\epsilon_\pi=8J-2 M_h$.}
This representation is shown in blue in fig.~\ref{fig:TFIM_correlator} (a). However, one can deform this branch cut to arrive at a representation which is more natural from a QFT perspective: the UV and the IR are now separated, \change{connecting $-2M_h$ with $+2M_h$ and the UV scale with infinity. The branch cuts along the real axis arise through the creation of fermion pairs with momenta $\pm p$ (i.e.\ zero net momentum), which act as branch points.}
There are extra poles coming from the Fermi-Dirac distribution function. These are nothing but (half of all) the  Matsubara frequencies $ \omega_n = - i \frac{\pi}{\beta} (2 n + 1)$ for integer $n$ and they are shown in red in  fig.\,\ref{fig:TFIM_correlator} (a). Their contribution is that of a transient effect. 

Let us stress here that so far we did not take the QFT limit ($\beta \, J \rightarrow \infty$ with $\beta \, M_{h}$ fixed) and, therefore, the presence of exponentially decaying terms in time is simply a feature of the spin chain in question. However, upon taking the QFT limit, they indeed do become exponentially decaying contributions known from QFT studies using CFT techniques and holography and motivating to a large degree the present work, see, e.g., refs.\,\cite{Birmingham:2001pj,Son:2002sd,Kovtun:2005ev}.

To make this point apparent, let us note that in (1+1)-dimensional CFT the retarded two-point function on a line at finite temperature is fixed by the conformal symmetry~\cite{Birmingham:2001pj,Sabella-Garnier:2019tsi}. For an operator of scaling dimension $\Delta$ and at $p = 0$, the retarded correlator has single poles at 
\begin{equation}
\label{eq.QNMsCFT}
\omega = - i\, 2 \pi\,T \, (\Delta + 2 \, n) \quad \mathrm{for} \quad n = 0,1,\ldots\
\end{equation}
which for $\Delta = 1$ coincide with single poles in eq.~\eqref{eq:TFIMcorrelator} originating from Matsubara frequencies. Indeed, for a canonically normalized operator of dimension $\Delta = 1$ the retarded CFT correlator up to contact terms reads
\begin{equation}
G_{R}^{{\cal O}_{\Delta = 1}}(t >0, p = 0) = -\frac{4 \pi }{\beta} e^{- \frac{2 \pi}{\beta} t} \left(1 - e^{- \frac{4 \pi}{\beta} t} \right)^{-1}
\end{equation}
and has a sequence of poles at positions given by eq.~\eqref{eq.QNMsCFT} with residues~$- \frac{4\pi}{\beta}$. One can check that eq.\,\eqref{eq:TFIMcorrelator} predicts the same residues for $M_{h} = 0$ when $\beta \, J \rightarrow \infty$. Therefore, for transients at $g = 0$ we can use the residue, \change{which, as detailed in the next section, we can identify with the coefficients accompanying the exponentials in the signal analysis,} to identify the QFT regime and we found the CFT regime reached for $\beta \, J$ between $5$ and $10$~\footnote{Cf.\ detailed analyses in appendix~\ref{app:C} of the supplemental material.}.

\section{Tensor networks setup and signal analysis}

In order to analyze the structure of singularities beyond exactly solvable cases, we evaluate numerically on the lattice the time and space dependent 2-point correlators of the form $\sim i \, \mathrm{tr}\left( \rho_{\beta} \left[\sigma^{\ell}_x(t),\sigma^0_x \right] \right)$.
This is achieved using standard TN techniques~\cite{Verstraete2008,Schollwoeck2011,Orus2014a}. We use finite systems with open boundary conditions and construct a MPO~\cite{verstraete2004mpo,Zwolak2004} approximation to the thermal state using the time evolved block decimation (TEBD) algorithm~\cite{Vidal2003} to simulate imaginary time evolution.
In this method, the evolution operator (in real or imaginary time) is written as a sequence of discrete time steps of width $\delta t$ (Trotter step), and each of them is approximated by a sequence of two-body operations, which are sequentially applied on the MPO ansatz. 
The application of these gates would generally increase the bond dimension of the MPO, thus a
\emph{truncation} is performed to keep it bounded.
The same method is later used to evolve the MPO in real-time after having perturbed it with a local operator, in order to produce the thermal response functions~\cite{barthel2009finiteT,karrasch2012finiteT}.
Thermal states of local Hamiltonians can be efficiently approximated by MPO~\cite{Hastings2006a,molnar2015thermal}, and the numerical error in this procedure is dominated by the truncation error induced by the real-time evolution. Its magnitude can be estimated by comparing the results obtained for different values of the maximal allowed bond dimension~$\chi$.

To reconstruct the analytic structure of the retarded correlator, we use Prony methods~\cite{peter2014generalized}, which represent the function as a sum of complex exponentials
\begin{equation}
\label{eq.prony}
    G(t) = \sum_{k=1}^N c_k e^{-i \, \Omega_k\, t},
\end{equation}
where $c_k$ and $\Omega_k$ are complex and $N$ is chosen. Such functions obey a linear differential equation and Prony methods exploit linearity to determine the frequencies $\Omega_k$ independently of the amplitudes $c_k$. While we specialized to the ESPRIT algorithm, there are many implementations. Some of them \change{are also known as \emph{linear prediction} and were used in the context of TN studies}~\cite{Pirvu_2010,barthel2009finiteT,Ganahl_2015,Steffens_2014,Wolf_2014,Wolf_2015}.

Although Prony only allows for discrete poles in frequency space, we can discover branch cuts, as they will be represented by sequences of poles. We will apply the Prony method to a sequence of time windows. We identify complex frequencies that are stable across different time windows with poles, while streaks are interpreted as branch cuts. To make it more apparent in our plots we use different colors to denote different time windows. The fuzzier a given structure is, the bigger its uncertainty~\footnote{Quantitative estimates of this uncertainty and further details are provided in appendix~\ref{app:C} of the supplemental material.}.

In order to probe the performance of the method, we apply it to the Ising CFT on a lattice. Fig.\,\ref{fig:TFIM_correlator}~(b) shows the analysis of eq.~\eqref{eq:TFIMcorrelator} when $M_{h} = 0$. Comparing with exact predictions one sees that the Prony analysis identifies both branch cuts and multiple decaying poles lying further away in the complex plane. This should be contrasted with the standard Fourier transform, used in a related context in, e.g., ref.\,\cite{kormos2017real}, which specializes in identifying poles that are close to the real axis.

As we have remarked before, the alignment of branch cuts and resulting pole structure is ambiguous, so Prony must implicitly make some choice here. In our experience, the method prefers to align branch cuts vertically, along the axis corresponding to the decay rate. Intuitively, such a choice is very efficient at late times because most contributions will be heavily suppressed.

In the next two sections we will use TN + Prony to predict the structure of singularities of correlators in interacting QFTs defined by the Hamiltonian eq.\,\eqref{eq:H_IsingQFT}.

\section{Singularities of the retarded thermal correlator: masses}

For a free fermion CFT, the branch cut along the real axis at $p = 0$ arises from the exchange of pairs of fermions of zero net momentum. As we already indicated, turning on a non-zero longitudinal field leads to a confinement of fermions and non-perturbative formation of bound states (mesons). Meson masses are known exactly in the integrable E$_8$ QFT and in other cases have been determined numerically using different methods.
\begin{figure}[!t]
\centering
 \includegraphics[width=1\linewidth]{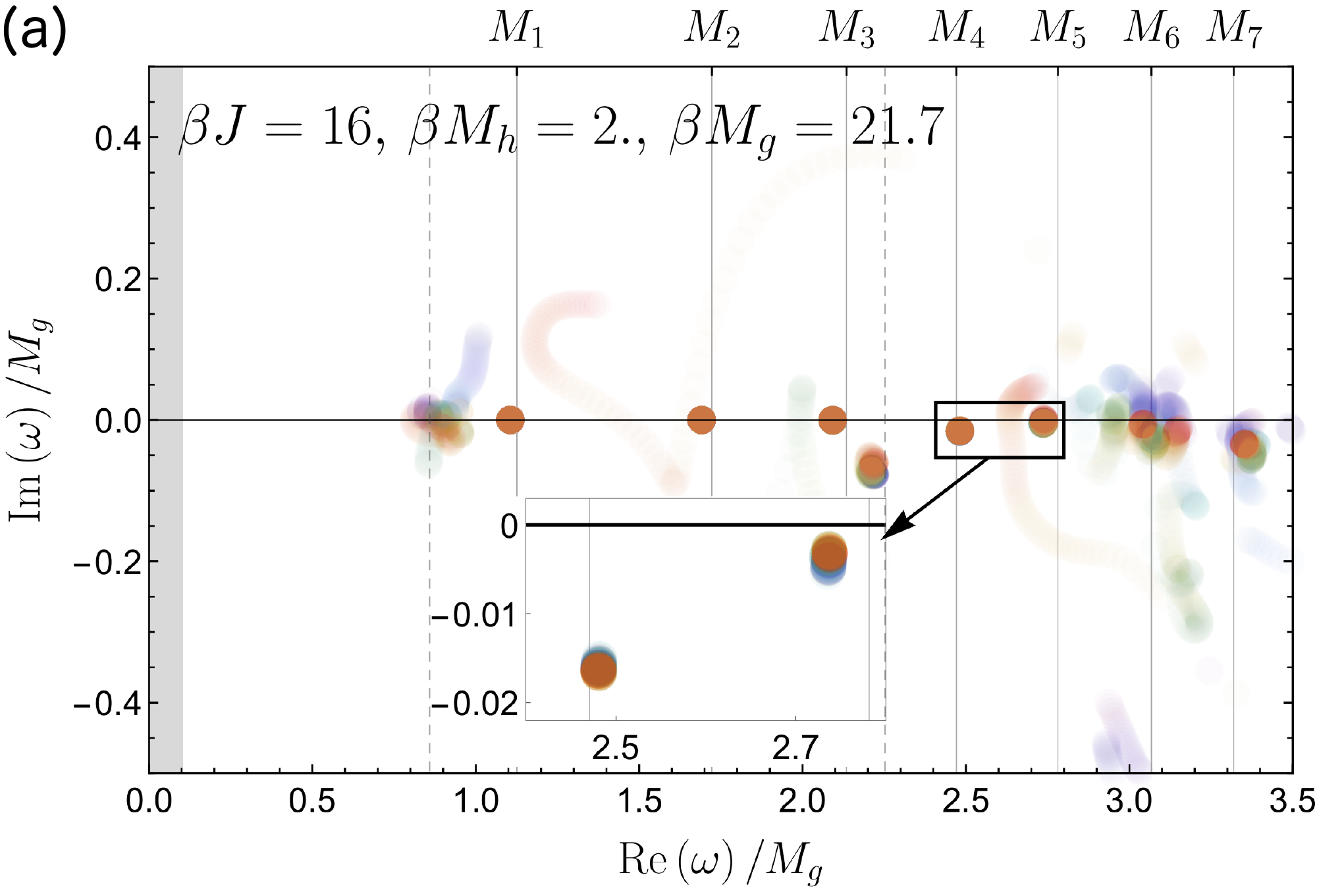} 
  \includegraphics[width=1\linewidth]{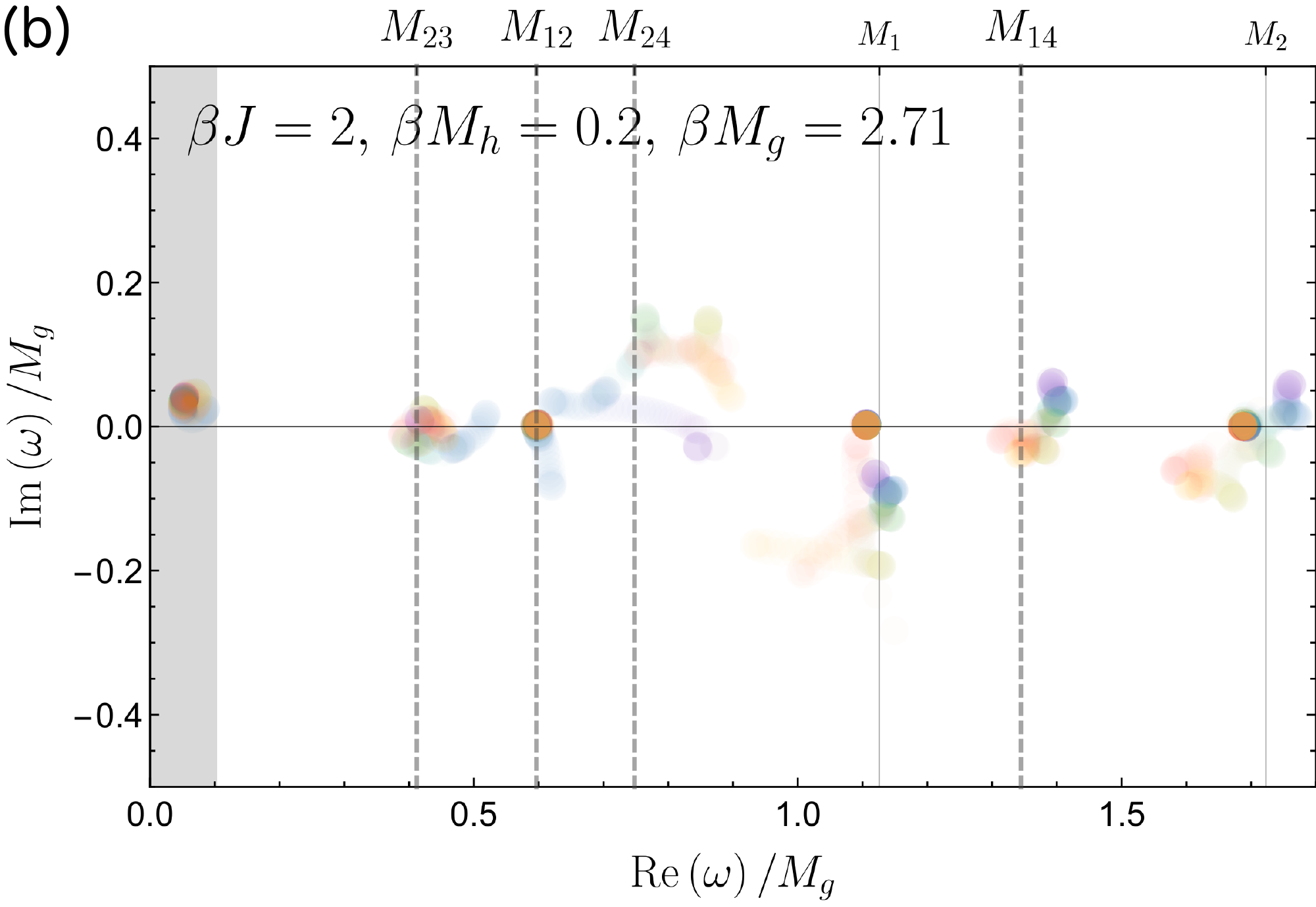}
 \caption{Prony analysis of the retarded correlator of $i \, \bar{\psi}\psi$ near the vacuum (a) and for the highest achieved temperature (b) in a non-integrable ferromagnetic case. The shaded region shows frequencies with wavelength not fitting in the signal window. The vertical lines indicating masses are taken from ref.~\cite{Fonseca:2006au}. (a): The dashed lines mark continuum threshold of $2 M_1$ and a boundary state calculated with DMRG. The inset zooms in on the $4^{\mathrm{th}}$ and the $5^{\mathrm{th}}$ meson to show their imaginary parts. All values are consistent with QFT. (b): The dashed lines indicate differences in masses appearing as a result of heating the system. Simulation parameters: $L=200, \chi = 170, J \, \delta t = 0.02,  J\,t_\text{max}=50$, $2^{\mathrm{nd}}$~order Trotter decomposition.}
 \label{fig:masses_nonintegrable}
\end{figure}
Our TN+Prony analysis shows that mesons enter the retarded correlator of $i\,\bar{\psi} \psi$ at non-zero temperature $\beta^{-1}$ and vanishing spatial momentum primarily as single poles corresponding to meson masses (and decay rate if unstable), see fig.~\ref{fig:masses_nonintegrable}.

Regarding the accuracy of our prediction, in table~ \ref{tab:e8_masses_ratio} we benchmark the results of a low temperature (basically vacuum) simulation extracted using TN+Prony with the analytic results for 7 out of 8 mesons existing in the E$_{8}$ QFT. In all the cases but the most massive meson we detect, our predictions match analytics within $1.5\%$ and in three cases within a fraction of percent~\footnote{See appendix~\ref{app:D} of the supplemental material for a discussion of quenches at zero temperature.}.
\begin{table}[b]
    \centering
     \begin{tabular}{l*{6}{c}} \hline
 & $M_2/M_1$ & $M_3/M_1$ & $ M_4/M_1$ & $ M_5/M_1$ & $M_6/M_1$ & $M_7/M_1$ \\ \hline 
 TN & 1.6147(7) & 1.962(1) & 2.413(2) & 2.936(3) & 3.165(6) & 3.52(3) \\
 E$_8$ & 1.6180 & 1.9890 & 2.4049 & 2.9563 & 3.2183 & 3.891 \\ \hline
\end{tabular}
    \caption{\label{tab:e8_masses_ratio} Ratios of masses of mesons extracted from TN+Prony compared to the analytical expectations for the integrable E$_8$ theory, e.g. $M_{2}/M_{1} = 2 \cos{\frac{\pi}{5}}$ \cite{Zamolodchikov:1989fp}. 
    }
\end{table}
We then consider interacting non-integrable case starting with a low temperature (with respect to the mass of the lightest meson $M_{1}$). The result from TN+Prony is shown in fig.~\ref{fig:masses_nonintegrable}~(a). As in the E$_8$ case, the frequencies agree with masses previously calculated in \cite{Fonseca:2006au, Zamolodchikov:2013ama}. However, some poles also develop an imaginary part, a signal that they are unstable. In particular, the ratio between the imaginary parts of the fourth and fifth mesons (see inset in fig.~\ref{fig:masses_nonintegrable}~(a)) from TN+Prony is $0.22 \pm 0.04$, which should be compared with the value obtained in ref.~\cite{Delfino2005Jul}, 0.233. This feature comes from the absence of integrability and a presence of continuum of states starting from the threshold of twice the mass of the lightest meson $M_{1}$.

\begin{figure}[!t]
\centering
 \includegraphics[width=\linewidth]{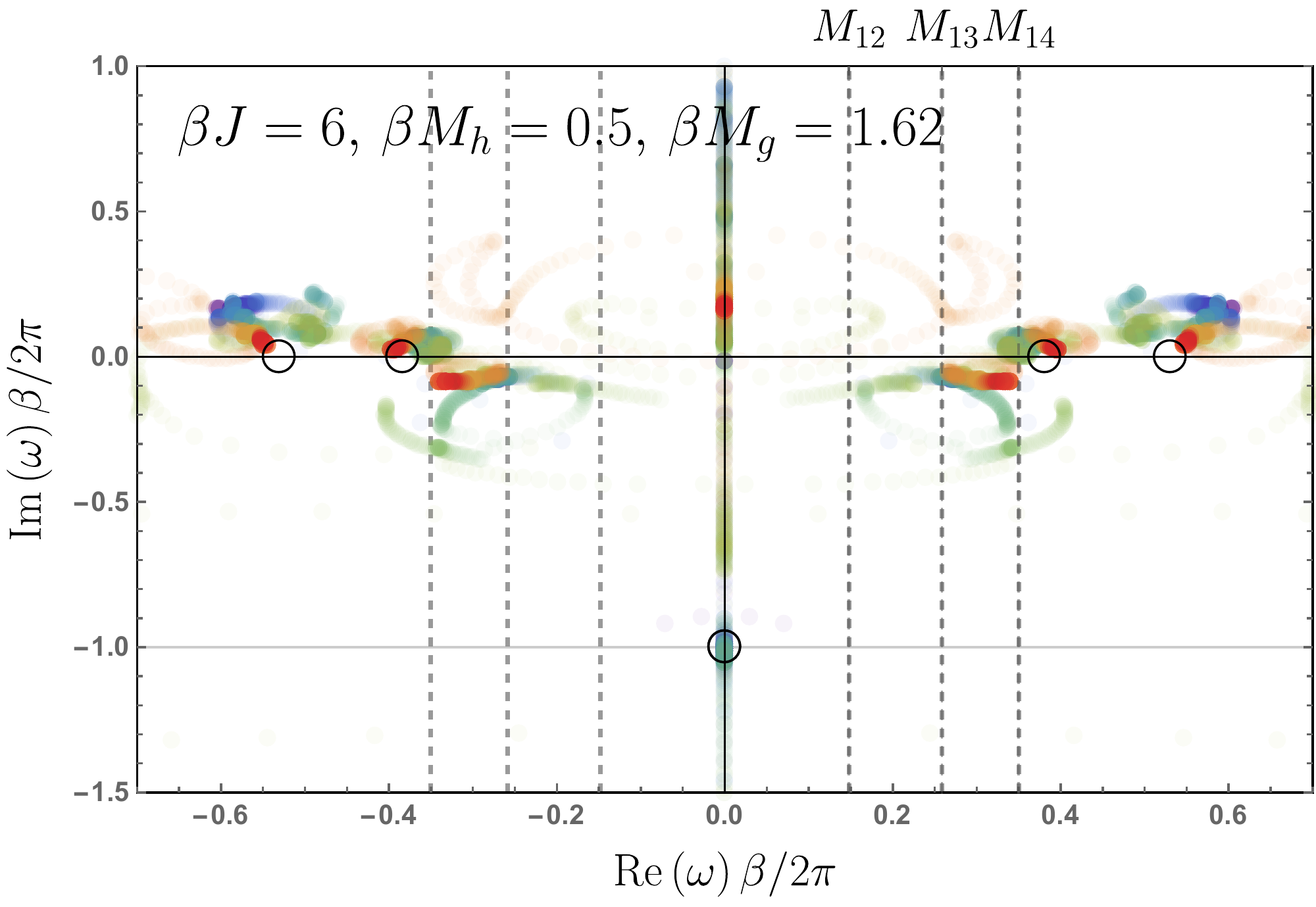}
 \includegraphics[width=.8\linewidth]{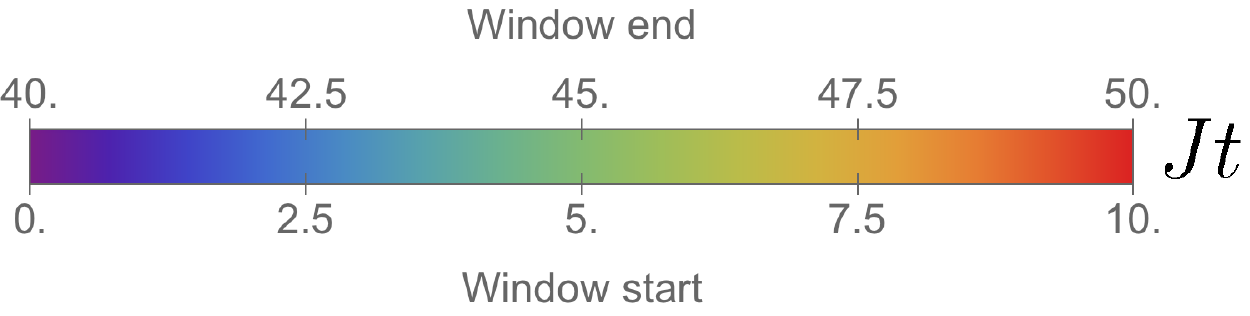}
 \caption{Prony reconstruction of the retarded correlator of $i\, \bar{\psi}\psi$ in the non-integrable ferromagnetic case for the continuum limit. The circles represent vacuum meson masses~\cite{Fonseca:2006au} and the leading transient for the free fermion QFT case, respectively. The numerical results are in reasonable agreement with these data. Dashed lines indicate mass differences. Poles on the positive imaginary axis are attributed to numerical artefacts. Simulation parameters: $L=200, \chi = 250, J \, \delta t = 0.005,  J\,t_\text{max}=50$, $2^{\mathrm{nd}}$ order Trotter decomposition.}
 \label{fig:mesons_and_decay}
\end{figure}

Apart from the mesons, there is a fuzzy structure appearing at a frequency slightly below $M_1$ as well as another pole with frequency slightly above $M_3$. This lower frequency arises due to open boundary conditions, which support excitations living close to the edges of the chain and is consistent with the result of a DMRG~\cite{White1992, Verstraete2004, Schollwoeck2011} calculation (dashed line). The higher frequency lies at $2 M_1$ (dashed line in fig.~\ref{fig:masses_nonintegrable}~(a)), which is the threshold of the two mesons continuum indexed by their relative momenta. This should be associated with a branch cut, i.e. a time dependent pole, and we believe this to be the source of the fuzziness.

Calculations are harder when the temperature is raised and become intractable in the regime where $\beta M_{1} \sim 1$. The clearest signature of thermal effects (understood as features of the correlator not present at very low temperatures) is the appearance of poles at locations corresponding to differences in masses, see fig.\,\ref{fig:masses_nonintegrable}~(b). Only $M_{12} = M_{1} - M_{2}$ appears as a clean pole, but there are signs of more. Up to our accuracy, we are not able to assess whether the mass or decay rates of mesons change with increasing temperature, but we can say that residues associated with these single poles in the correlators decrease. In practice, what we mean by this is that the corresponding coefficients $c_{k}$ in eq.~\eqref{eq.prony} decay with temperature at fixed values of $M_{h} / M_{g}$ and $M_{g} / J$. It is in line with the expectation that at sufficiently high temperatures the natural degrees of freedom become fermions.

\section{Singularities of the retarded thermal correlator: the leading transient}

\begin{figure}[!t]
\centering
 \includegraphics[width=0.49\textwidth]{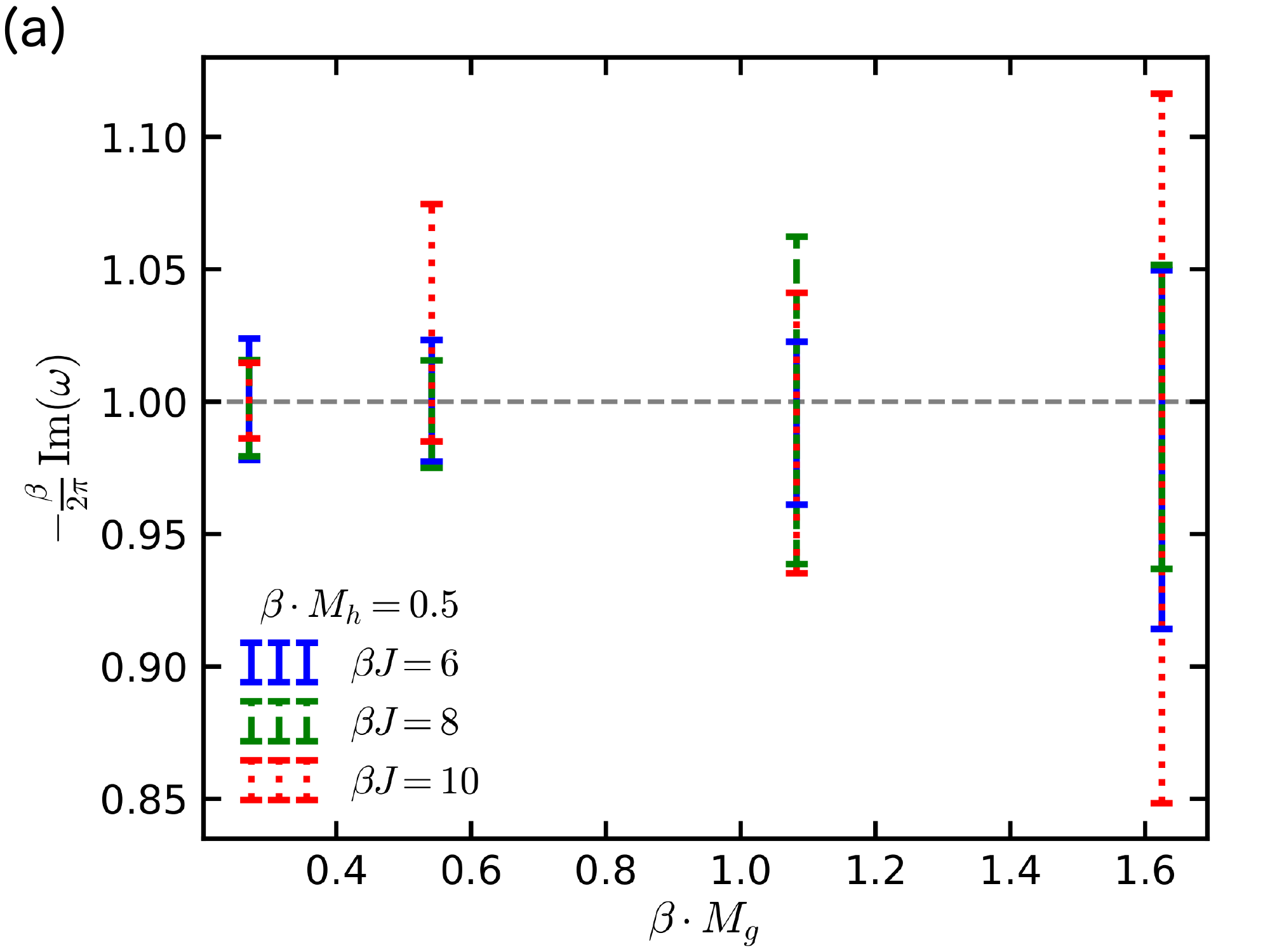}
 \includegraphics[width=0.49\textwidth]{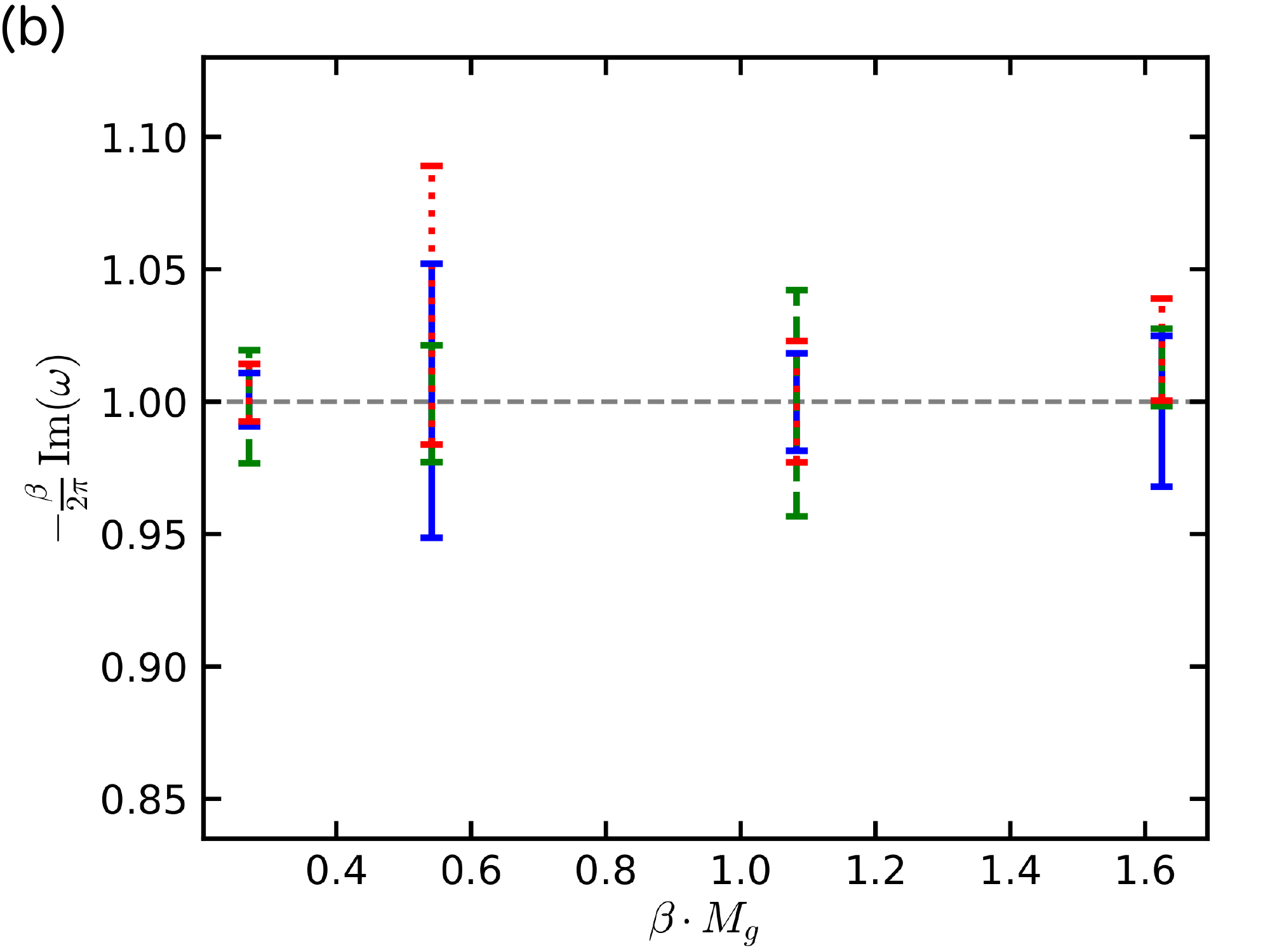} 
 \caption{Extraction of the least damped transient pole for the non-integrable case using TN+Prony. The continuum limit is approached (by increasing $\beta J$) in the ferromagnetic (a) and paramagnetic phase (b) for fixed $\beta M_h = 0.5$ and for different values of $\beta M_g$.}
 \label{fig:cont_limit_nonint_TDpoles}
\end{figure}

Beyond meson signatures, an intrinsically thermal feature of the linear response theory are single poles in the complex frequency plane that lead to features of the correlator decaying exponentially in time (transients). In fig.~\ref{fig:TFIM_correlator} (b) one sees indications that the CFT prediction for the two least damped contributions to the retarded correlator of $i\,\bar{\psi} \psi$ is captured by the Prony method applied to eq.~\eqref{eq:TFIMcorrelator}. Similar finding also holds when one applies TN to the same setup.

Based on this observation, we investigate whether it is possible to simultaneously identify the leading transient and meson states through Prony analyses. 
Such an non-integrable example is shown in fig.~\ref{fig:mesons_and_decay}. The decaying pole is naturally most visible at early time windows in the Prony analyses, while the first two meson states are best visible at late times.
This demonstrates that the correlator signal contains proper information of the QFT regime (through the existence of non-perturbative QFT bound states) and thermal effects (decaying poles).

The question that we want to systematically address now is what happens to singularities of the retarded correlator residing in free QFTs at locations specified by eq.~\eqref{eq.QNMsCFT} when we turn on interactions, i.e. for non-zero~$g$~(or, equivalently, $M_{g} \neq 0$). To shed light on this issue, we fix the transverse field by $\beta M_h = 0.5$ and investigate the behaviour as the longitudinal field is increased as $\beta M_g \approx \{ 0.27,  0.54,  1.08,  1.62 \}$.
The continuum limit is approached by increasing $\beta J=  6,\,8,\, \mathrm{and} \, 10$.
TN simulations of the correlator are then performed for a spin chain of size $L=100$ in a time interval up to $J \, t_{max}=10$.

What we observe is that the single pole singularity governing the leading exponential decay not only survives in the presence of interactions, but, quite surprisingly, for the value of parameters we consider we do not observe significant systematic deviations from the free fermion QFT result $\frac{1}{2\pi}\operatorname{Im}(\beta \, \omega) = -1$, see fig.~\ref{fig:cont_limit_nonint_TDpoles}~\footnote{We refer to the supplemental material for details on the MPO simulations (appendix~\ref{app:B}) and the uncertainty estimation based on the Prony analyses (appendix~\ref{app:C}).}. 
This conclusion holds both for the ferromagnetic as well the paramagnetic phase. 
In particular, when increasing the interaction scale $M_{g}$ with respect to both $\beta$ and $M_{h}$, and also when increasing $\beta\, J$, the uncertainty in determining the value of the corresponding complex frequency grows up to 13\,\% in the ferromagnetic and 5\% in the paramagnetic phase.

\section{Summary}

Although real-time simulations with TN are usually reliable only for a finite time window, we have shown here that combining moderate time TN runs
with a Prony analysis allows us to extract
the analytic structure of the retarded thermal two-point function in the complex frequency plane, which provides an invaluable insight into dynamics.
\change{Notice that our method does not try to reconstruct the full details of time evolved correlators beyond the time window in which simulations are reliable, but utilizes these data to extract relevant features of the model, to identify the frequency poles.}
By suitably choosing parameter regimes, this strategy enables us to make ab initio predictions about the response of thermal states to perturbations,
for a class of non-integrable interacting~QFTs. 

We have probed the accuracy of the method in the integrable cases of the free fermionic Ising and the interacting E8 QFTs. Also for non-integrable QFTs we reproduced
both the real and imaginary parts of several mesons frequencies that were earlier found in
refs.~\cite{Zamolodchikov:1989fp, Fonseca:2006au, Zamolodchikov:2013ama, Delfino2005Jul}. For increasing temperatures, we find that residues of meson poles decrease, and poles start appearing at frequencies corresponding to differences in masses. Our results are stable with respect to variations in the system size, bond dimension and other numerical parameters.

Another effect triggered by the temperature are transient contributions to the correlator, which in the free fermion case can be related to Matsubara frequencies~\cite{Kurkela:2017xis} and for holographic CFTs correspond to quasinormal modes of anti-de Sitter black holes~\cite{Son:2002sd,Kovtun:2005ev}. 
We have also extracted these numerically and within our numerical precision and in the explored regime, these transients are not affected by interactions or integrability breaking.

\begin{acknowledgments}
	We would like to thank Enrico Brehm, Pasquale Calabrese, Pawel Caputa, Jens Eisert, Lena Funcke, Philipp Hauke, Romuald Janik, Aleksi Kurkela, Stefan Kuehn, Mikko Laine, Giuseppe Mussardo, Rob Myers, Milosz Panfil, Volker Schomerus, Sukhbinder Singh, Gabor Takacs and Stefan Theisen for useful discussions and correspondence. The Gravity, Quantum Fields and Information group at AEI is generously supported by the Alexander von Humboldt Foundation and the Federal Ministry for Education and Research through the Sofja Kovalevskaja Award.
	This work was partly supported by the Deutsche Forschungsgemeinschaft (DFG, German Research Foundation) under Germany's Excellence Strategy -- EXC-2111 -- 390814868, and by the EU-QUANTERA project QTFLAG (BMBF grant No. 13N14780).
JK and VS are partially supported by the International Max Planck Research School for Mathematical and Physical Aspects of Gravitation, Cosmology and Quantum Field Theory.
The work of JK is supported in part by a fellowship from the Studienstiftung des deutschen Volkes (German Academic Scholarship Foundation). Some of our calculations were performed using the excellent tensor network package TeNPy~\cite{tenpy}.
\end{acknowledgments}

\bibliographystyle{jk_ref_layout_wTitle} 
\bibliography{literature}

\clearpage
\appendix

\section{Jordan-Wigner transformation and QFT limit\label{app:A}}

It is well-known that one can define complex fermionic operators $b_{j}$ via the Jordan-Wigner transformation~\cite{JordanWigner1928}, in terms of which 
\small
\begin{equation}
\sigma_x^{j} =1-2 \, b_j^{\dagger}\, b_j\quad \mathrm{and} \quad \sigma_z^{j} =\left(\prod_{l<j} 
(1-2 \, b_l^{\dagger}\, b_l)\right)(b_j+b_j^{\dagger}).
\label{eq:JW}
\end{equation}
\normalsize
When the longitudinal field vanishes ($g = 0$),
eq.~\eqref{eq:Isingham} reduces to a free fermion Hamiltonian.

We can now define two independent Majorana fermion fields as
$\psi(x=ja)=\sqrt{\pi/a} (b_j^{\dagger}+b_j) $, 
and $\bar{\psi}(x=ja)=-i\sqrt{\pi/a}(b_j^{\dagger}-b_j) $, 
where we have introduced a lattice spacing $a$ that we take to be $a=2/J$, which implies that the speed of light is 1.
With this prescription, in the continuum limit $a\to 0$ the fields anti-commute with each other and, on top of this, satisfy $\{\psi(x),\psi(y)\}= \{\bar{\psi}(x),\bar{\psi}(y)\} = 2 \pi\, \delta(x-y)$. The latter conditions ensure that their two-point correlation functions in the vacuum in the conformal field theory (CFT) regime (see below) decay precisely as $\frac{1}{x-y}$ at long distances.

\section{Generalities about the MPO simulations\label{app:B}}

\begin{figure}[!t]
\centering
 \includegraphics[width=0.49\textwidth]{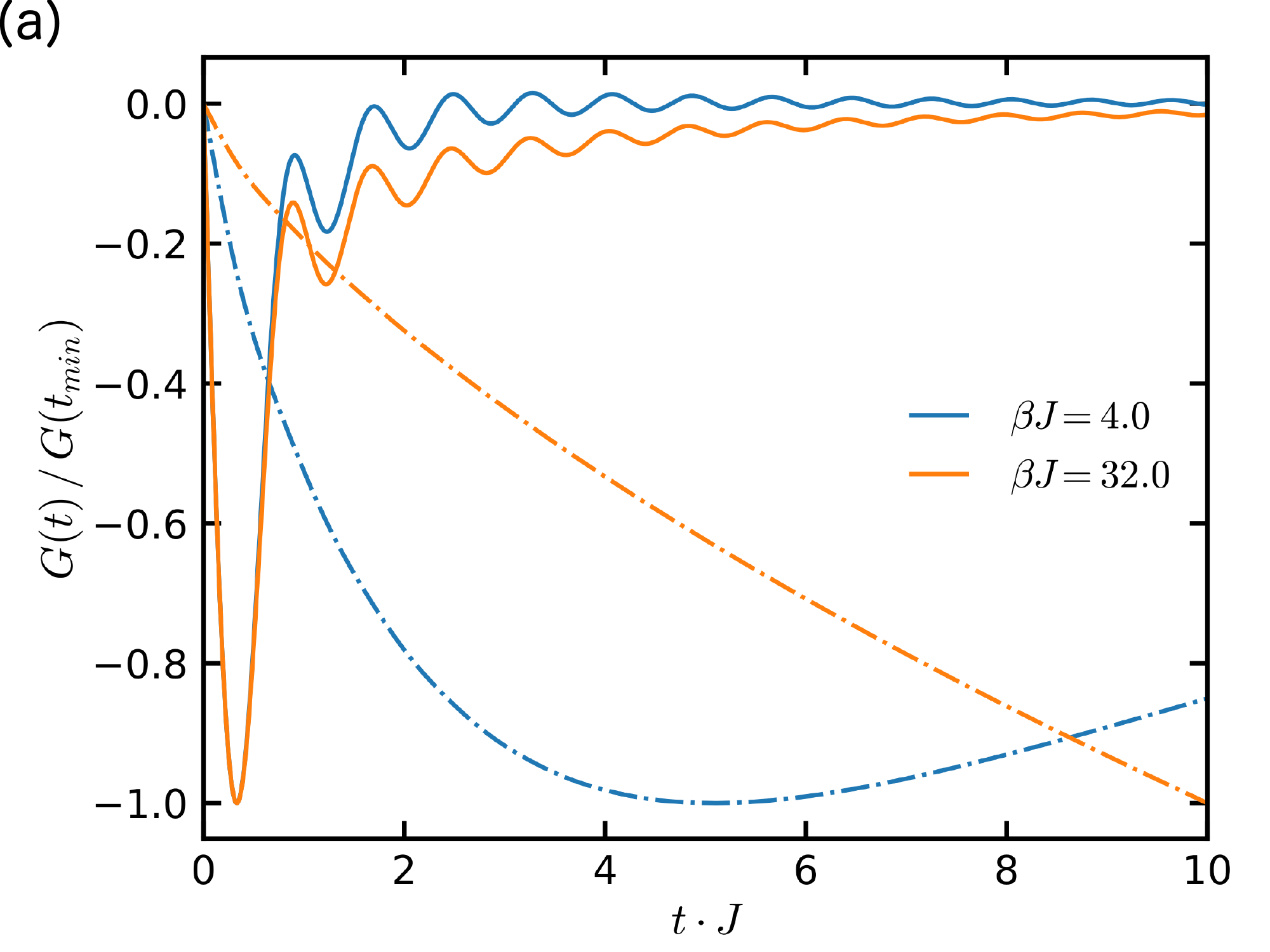}
 \includegraphics[width=0.49\textwidth]{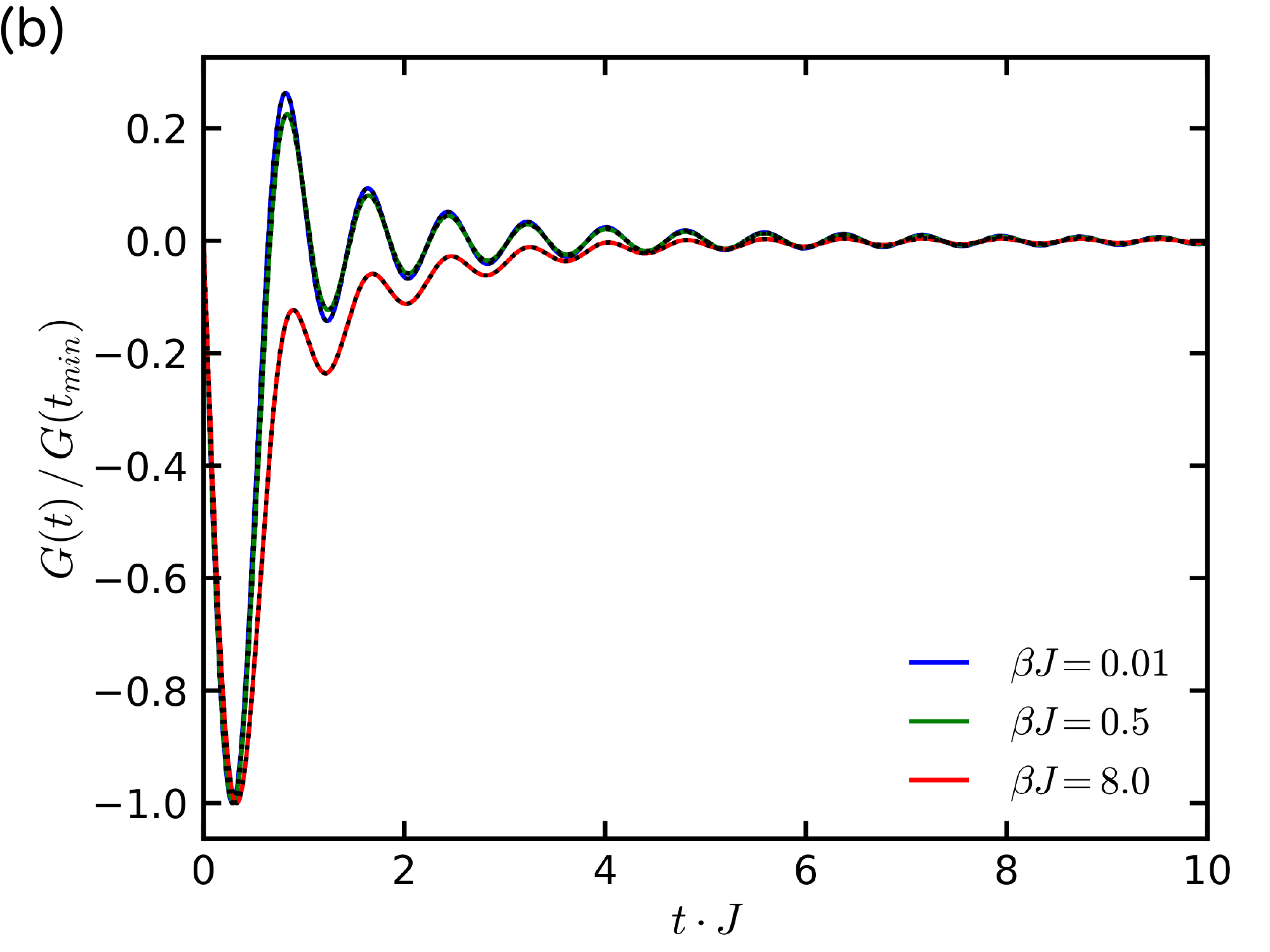} 
 \vspace{-4mm}
 \caption{(a): Comparison of the transverse (solid curves) and longitudinal (dash-dotted curves) response function obtained with MPO simulations (for $\beta M_h=0.2$, $\beta M_g =0$ in the ferromagnetic phase) and two values of the lattice spacing corresponding to $\beta J=4$ and $\beta J=32$. Note that the longitudinal response is oscillating on a much longer time scale.\\
 (b): Comparison between the MPO simulations (solid lines) and the exact results from free fermions (black dotted lines) for the transverse response function at the critical point ($\beta M_h = \beta M_g = 0$). We observe an excellent agreement for all the lattice spacing values $\beta J$. \\
  Numerical parameters: $L=100, \chi = 200, J \delta t = 0.005,  J t_\text{max}=10$, $2^\mathrm{nd}$ order Trotter decomposition.
 Both plots are scaled w.r.t.\ the minimum of the correlation function for visual purposes.}
 \label{fig:trans_vs_long}
\end{figure}

The central quantity in our studies is the retarded 2-point function at non-zero temperature.
This linear response function can be calculated w.r.t.\ the longitudinal or transverse magnetization. 
As discussed in the main text, the scaling dimensions of the corresponding relevant CFT operators differ by a factor of 8. 
This means that the decay and oscillation rate for the longitudinal response is too slow to observe within a time scale which our simulations are able to cover.
This applies also to the UV frequencies as visible in fig.\,\ref{fig:trans_vs_long} (a), in which the longitudinal response oscillates with a much longer time period.

In the absence of longitudinal field, we can exploit the mapping to free fermions in order to compare the MPO simulations to the analytical results.
Fig.\,\ref{fig:trans_vs_long} (b) shows the (scaled) transverse response function at criticality for a large range of temperatures.
The MPO simulations (colored solid curves) are superimposed with the integrable solution (dash-dotted curves), demonstrating very good agreement over the whole time range.
Near the ground state for $\beta J=8$, the signal shows some qualitative variation compared to higher temperatures.
This is also the regime which is relevant for taking the continuum limit $\beta J \to \infty$, which is described in detail in the next section.

\section{ The quantum field theory limit\label{app:C}}

\subsection{The integrable free fermion case}

\begin{figure*}[!t]
\centering
 \includegraphics[width=0.49\textwidth]{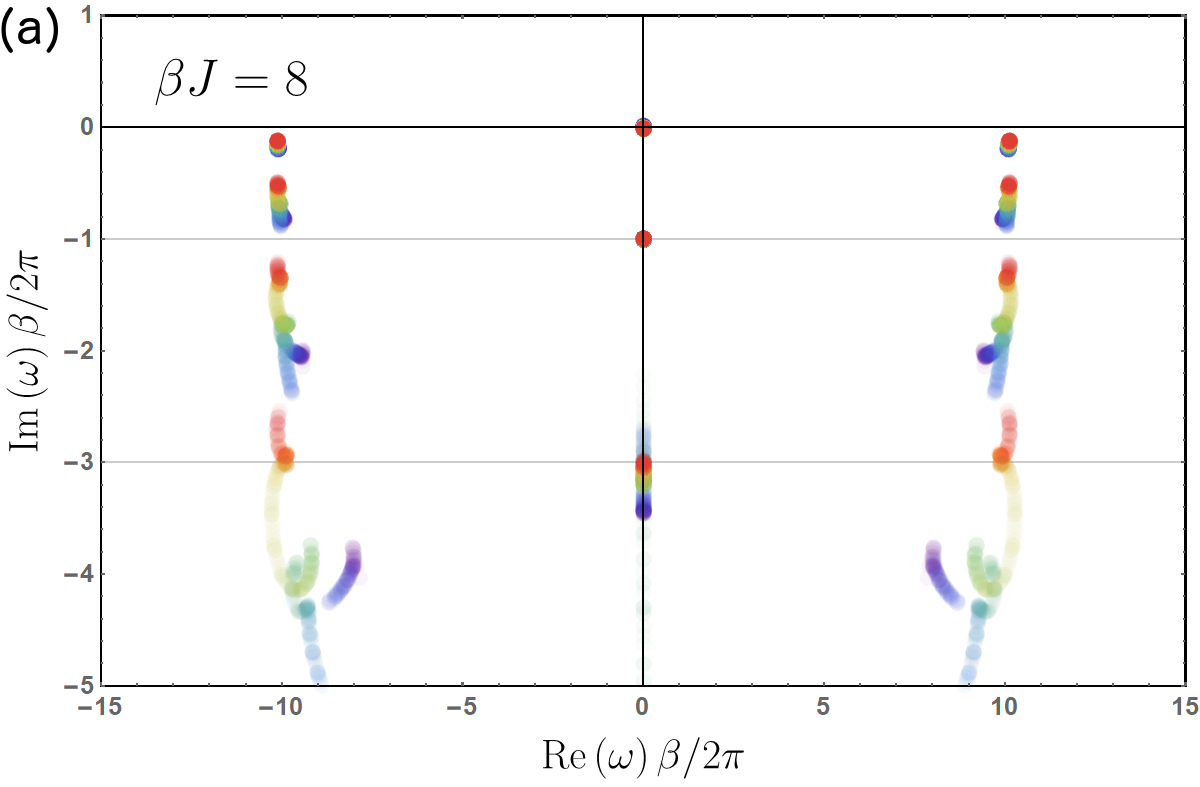} 
 \includegraphics[width=0.49\textwidth]{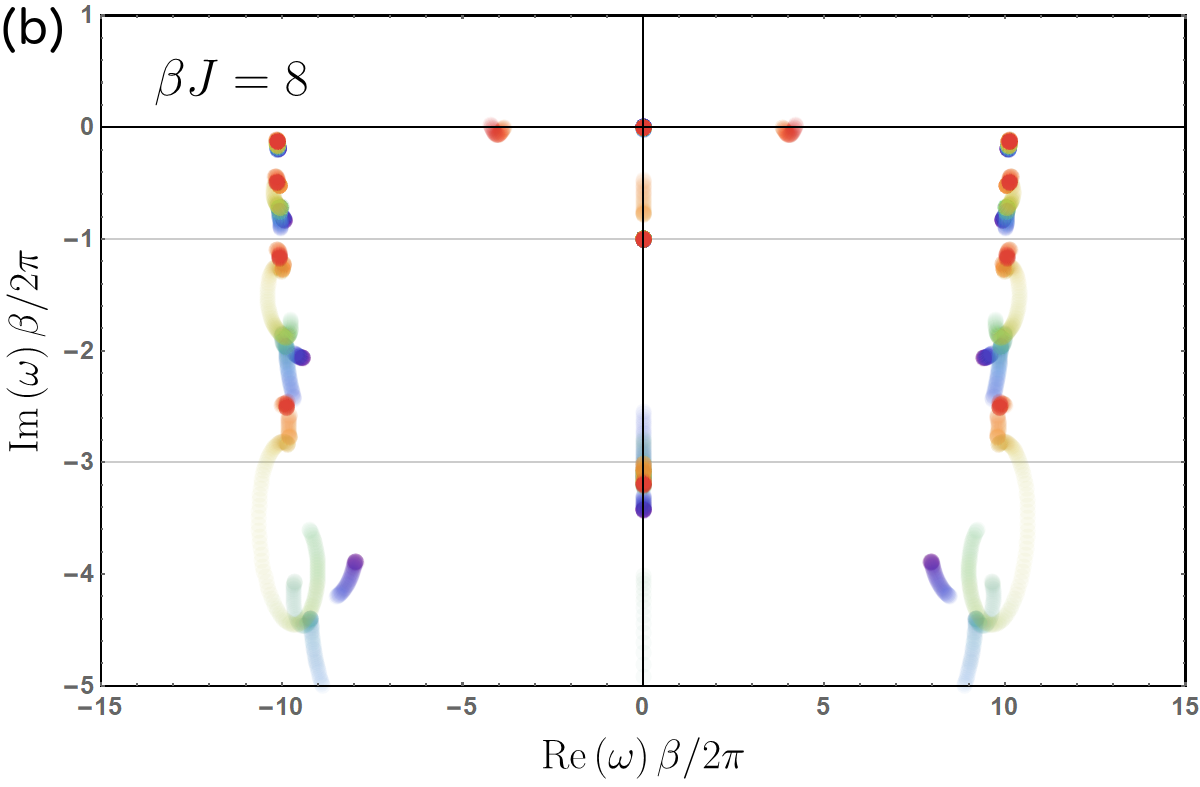}\\
 \includegraphics[width=0.49\textwidth]{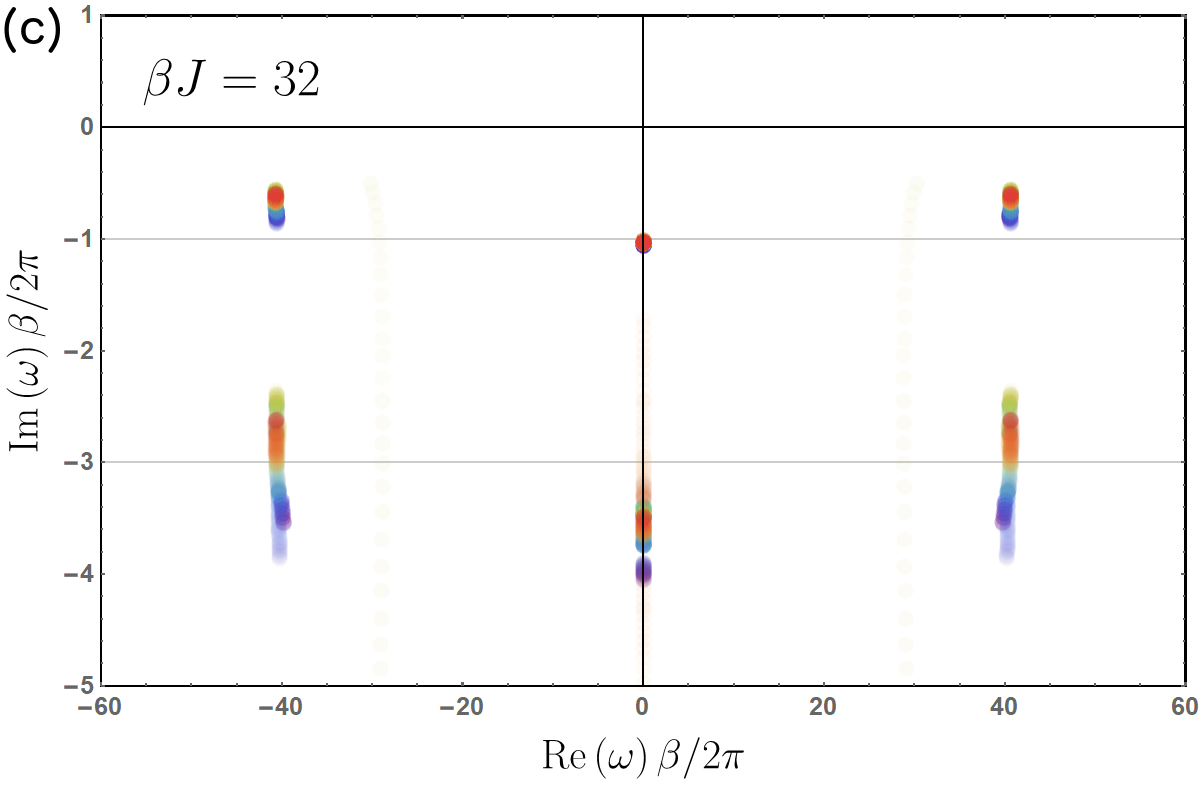} 
 \includegraphics[width=0.49\textwidth]{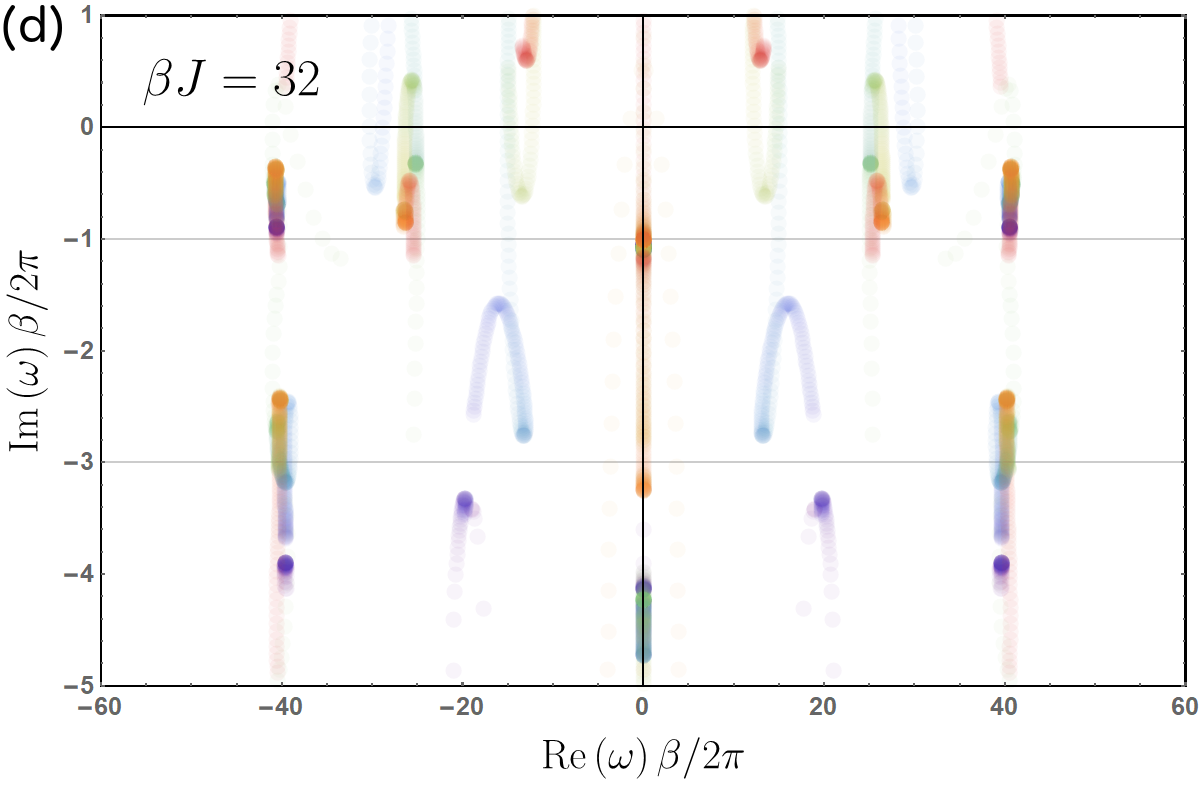} \\
 \includegraphics[width=0.4\textwidth]{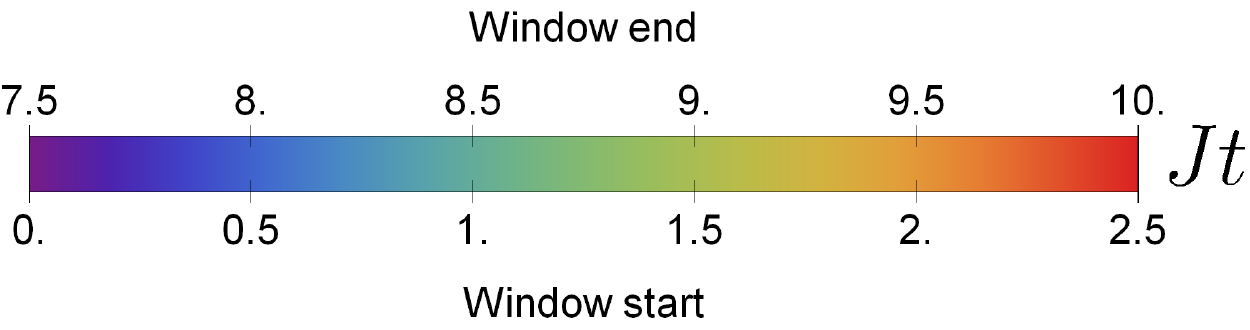} 
 \caption{Frequency analysis from Prony for the continuum limit in the integrable case. The figures show the complex values $\omega/J$ found by the Prony analysis in a particular time window as indicated by the color bar.
 The left column (a,c) is based on the analytical result for the response function and the right column (b,d) is obtained with MPO simulations. The imaginary axis is scaled by $\beta /(2\pi)$, such that the first two transient poles are located at $-1$ and $-3$.\\
 Numerical parameters: $L=100, \chi = 200, J \delta t = 0.005,  J t_\text{max}=10$, $2^\mathrm{nd}$ order Trotter decomposition.}
 \label{fig:cont_limit_g0}
\end{figure*}

When $M_g = 0$, the Hamiltonian can be exactly solved by mapping to free fermions as sketched in the main text. We can thus apply the Prony analysis to the exact results and compare to the TN numerics.

We set $\beta M_h = 0.2$ and approach the continuum limit by increasing $\beta J$, using the values $\beta J=\{2,4,8,12,16,32\}$. Going to the continuum limit means that the IR length scale gets larger and we must make sure that it does not exceed the spin chain size. For a chain of $L=100$ sites, the length scale of the mass compared to the chain is $J/(L M_h)$, which may be too large for $\beta J > 12$. 
The sequence of $\beta J$ corresponds to a series of increasing values of the transverse field $h=\{0.95,0.975,0.9875,0.991667,0.99375,0.996875\}$, i.e. approaching the critical point at $h=1$ from the ferromagnetic phase.

In fig.\,\ref{fig:cont_limit_g0} the result of the Prony analysis is shown for two selected values of $\beta J$. The left column in fig.\,\ref{fig:cont_limit_g0} is based on the analytical result for the infinite system while the right column is based on MPO simulations for a $L=100$ spin chain. From the analytical structure of the correlator, we know that there is a branch cut stretching between the two points $\pm 2 M_h$ (near the real axis) and another pair of cuts starting at $8J-2 M_h$ stretching out to infinity.
These figures demonstrate how the latter branch cuts are approximated by Prony analysis through a nearly vertical line of poles in the lower imaginary plane. 
Furthermore, there are purely decaying transient poles on the imaginary axis, located at $-\frac{2\pi}{\beta J}(2n+1)$ for $n = 0, 1, 2, \ldots$ \cite{Birmingham:2001pj}.
In fig.\,\ref{fig:cont_limit_g0}, these transient poles would be located at $-1,-3,\ldots$ on the rescaled imaginary axis.
The first transient pole is for both values of $\beta J$ clearly visible in the analytical result as well as in the MPO simulation.
For $\beta J=8$ also the second pole is visible in both results, while at $\beta J=32$ only in the analytical case.
Overall, the analytic and tensor network picture agree very well.
Since the values of the masses are small compared to the time windows, the branch cut between $\pm 2 M_h$ cannot be clearly resolved. At $\beta J = 32$, one would expect strong finite size effects for a $L=100$ chain and indeed, the Prony result contains many spurious and erratic poles.

\begin{figure}[!h]
\centering
 \includegraphics[width=0.49\textwidth]{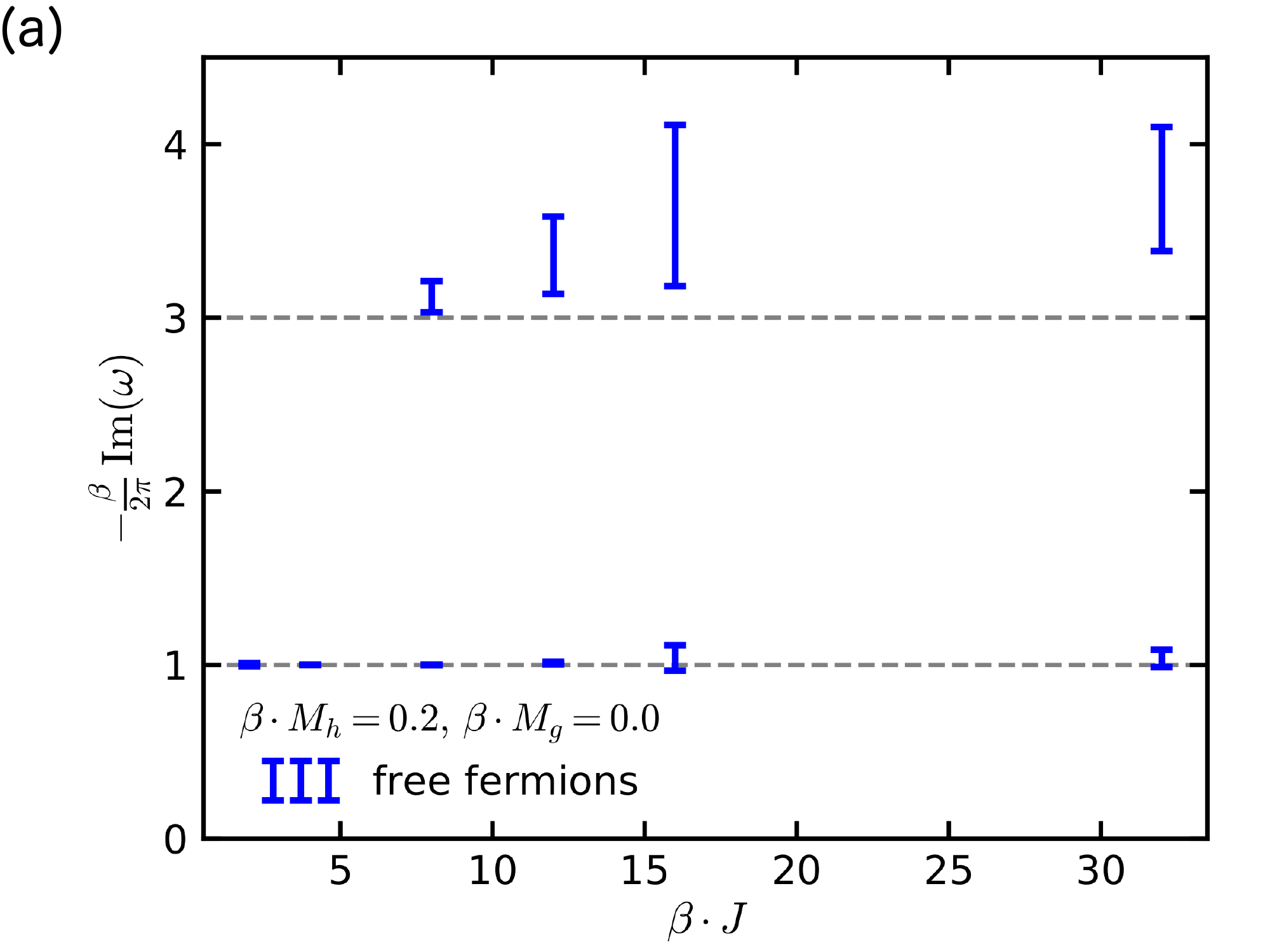}
 \includegraphics[width=0.49\textwidth]{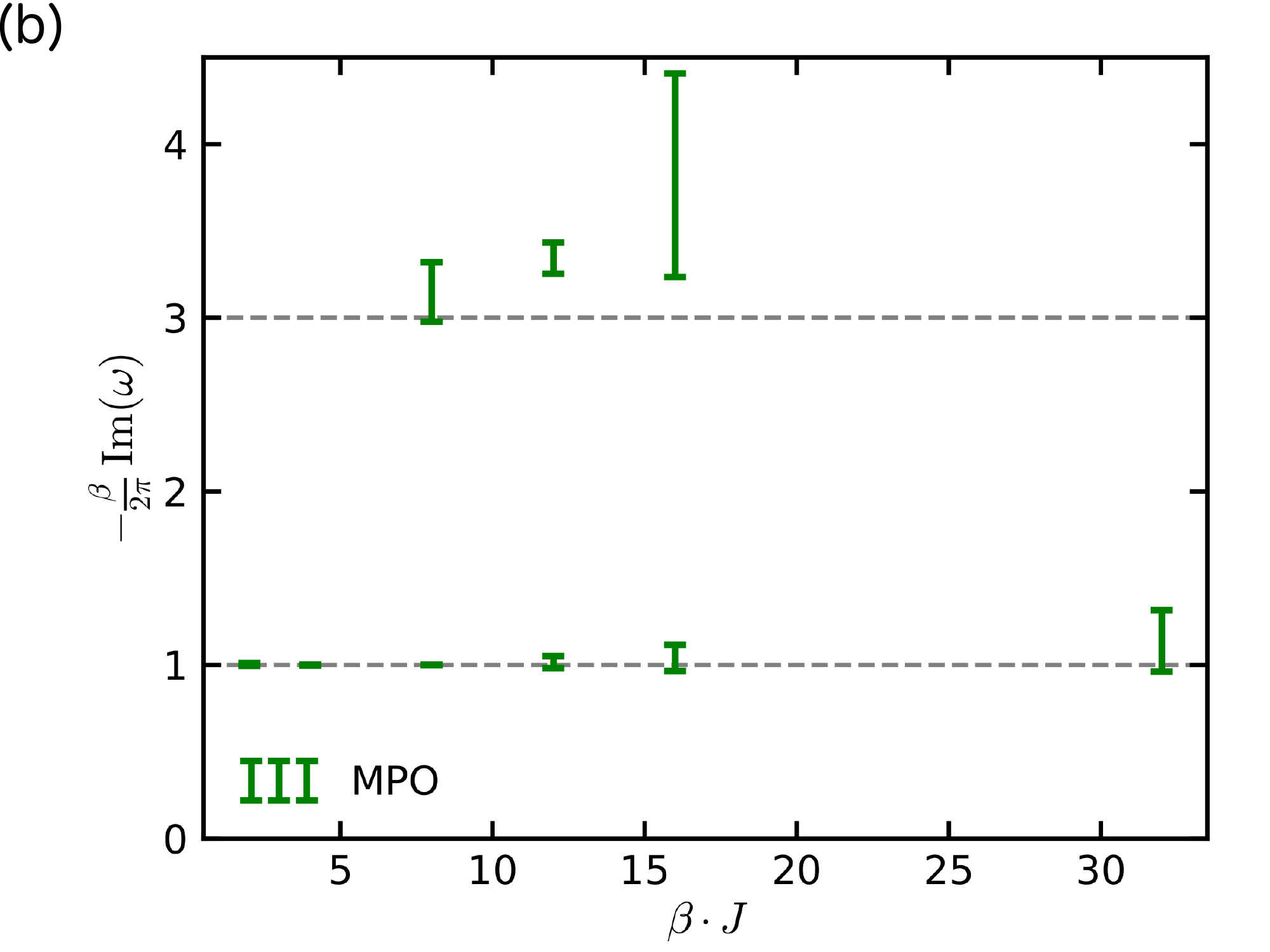} 
 \caption{Extraction of the purely decaying transient poles for the integrable case, based on Prony applied to analytical result (a) and MPO simulation (b). Error bars denote the calculated location including an uncertainty measure, see the detailed discussion in the text. The grey lines represent the correct result. As $\beta J$ increases, one is approaching the continuum limit and there are two explanations for why the results start to deviate. Either the chain is too small or the time window is too short. Since also the free fermion calculation suffers (which uses an infinite chain), it must be the time window.}
 \label{fig:cont_limit_g0_TDpoles}
\end{figure}

\begin{figure}[!t]
\centering
 \includegraphics[width=0.49\textwidth]{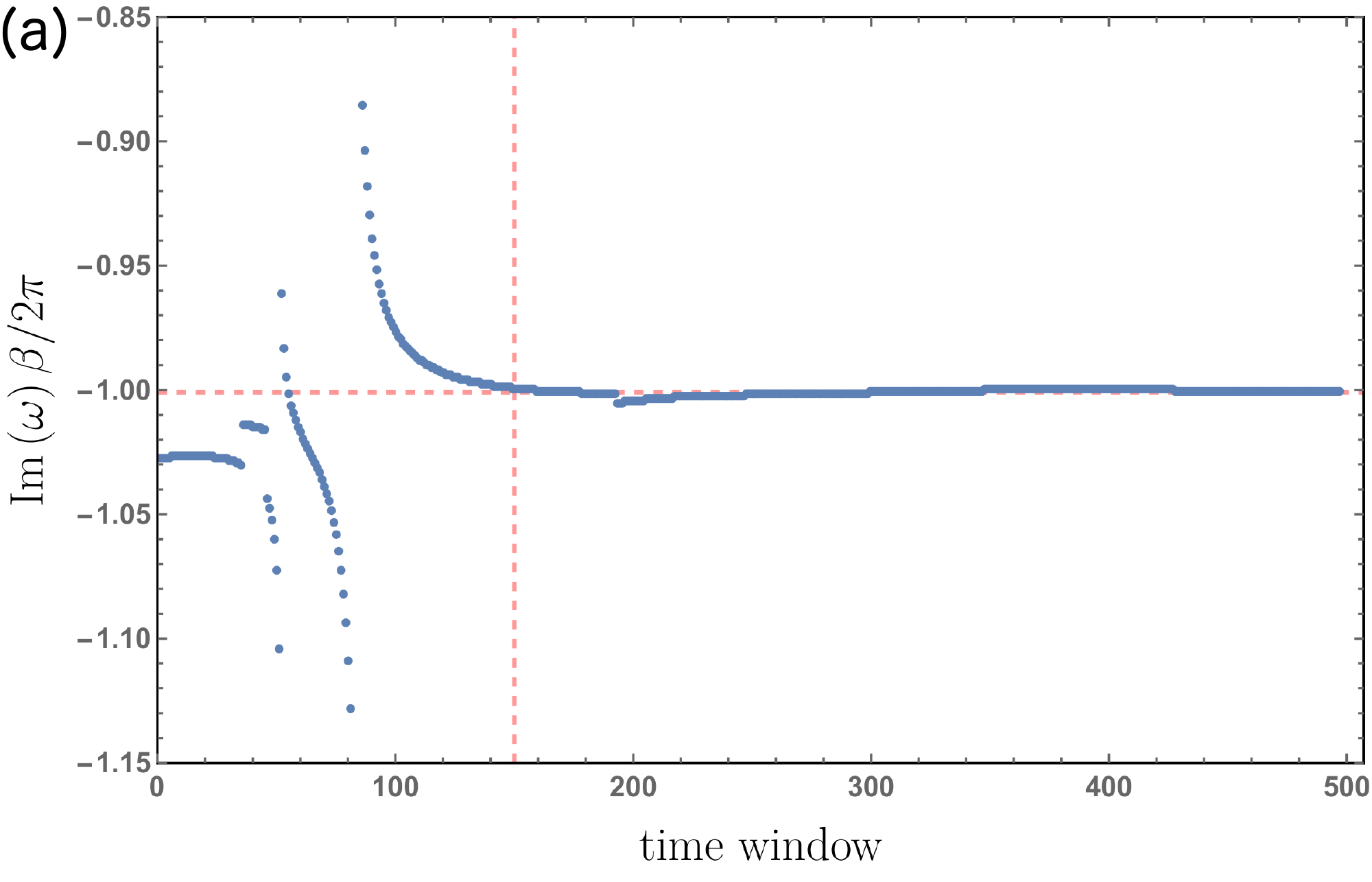}
 \includegraphics[width=0.49\textwidth]{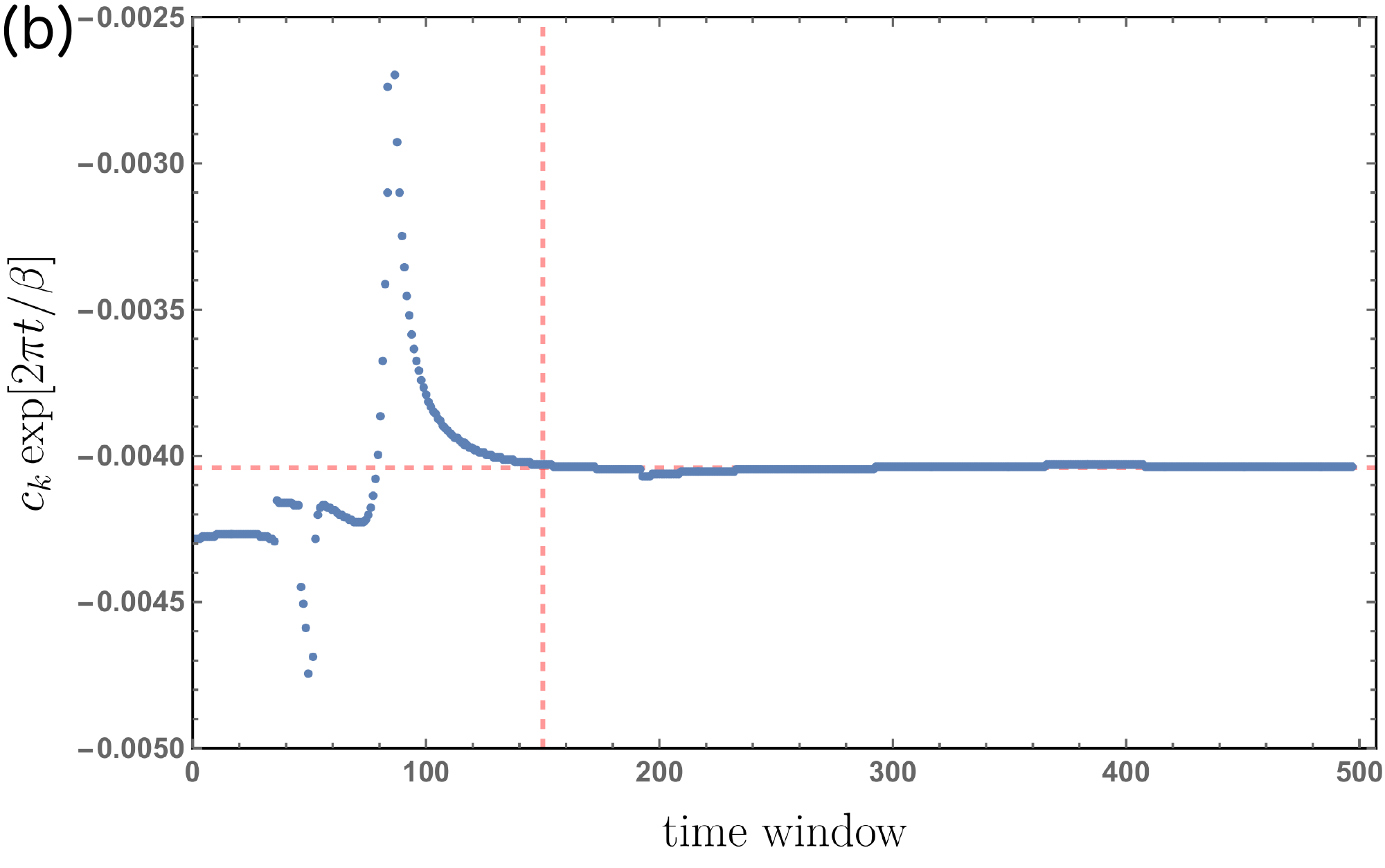} 
 \caption{Example of the Prony analysis for the MPO simulation of the retarded correlator at $\beta J=8$. Plot (a) shows the position of the first transient on the imaginary axis for every time window (with time increment step $\delta t=0.005$). Panel (b) plots the corresponding coefficient, multiplied with the inverse time dependence $\exp[2\pi t / \beta]$. For time windows later than the vertical red line indicates, the average value shown by the horizontal red line was calculated.}
 \label{fig:PoleAnalyses}
\end{figure}

Fig.\,\ref{fig:cont_limit_g0_TDpoles} shows how the position of the transient poles is consistent with the expectation as we approach the continuum limit by increasing $\beta J$.
The results from Prony analysis based on the free fermion calculation (blue, (a)) and MPO simulation (green, (b)), is represented by the error bars. These should be compared to the analytical values for the first and second transients: $-\frac{1}{2\pi}\operatorname{Re}(\beta \omega) = 1$ and $-\frac{1}{2\pi}\operatorname{Re}(\beta\omega) = 3$.
For this analysis, we have applied the Prony method on $75\%$ of all discrete points in the time evolution and then successively shifted the analysis window towards later times.
Based on the statistics obtained in this way, we calculated the mean value and standard deviation of all poles around the analytical positions.
A selected example of such an analysis is shown in fig.\,\ref{fig:PoleAnalyses} (a) for the first transient.
Therein, the location of the first decaying pole on the imaginary axis is plotted for every time window. At early times, the identification is rather unstable, which is presumably related to the fact that higher order transients are not yet decayed and overlapping the signal.
We thus choose a late enough time window to calculate the mean value and deviation of the pole (indicated by the red lines).
The resulting uncertainty of this analyses is plotted in fig.\,\ref{fig:cont_limit_g0_TDpoles}.
The first transient pole can be identified in the whole temperature or mass range. 
For $\beta J \le 8$ the pole can be resolved with an accuracy below $1\%$.
For $\beta J=32$ the uncertainty increases to $5\%$ (free fermions) or $18\%$ (MPO).
The values based on the analytical result and MPO simulation differ otherwise only marginally.

The second pole can be only identified in the temperature (or, equivalently, lattice spacings) range $\beta J=8 \ldots 32$ (analytical case) or $\beta J=8 \ldots 16$ (MPO simulations) with large uncertainty.
The reason for this behavior is the fact that at low $\beta J$ the branch cut stretching from the UV obscures the poles on the imaginary axis.
On the other hand, at large $\beta J$, the correlation length increases such that finite size and finite time effects become important in MPO simulations. Since the analytic results agree with the MPO results, it is finite time rather than finite size that is responsible for the deviations.

\begin{figure}[!h]
\centering
 \includegraphics[width=0.49\textwidth]{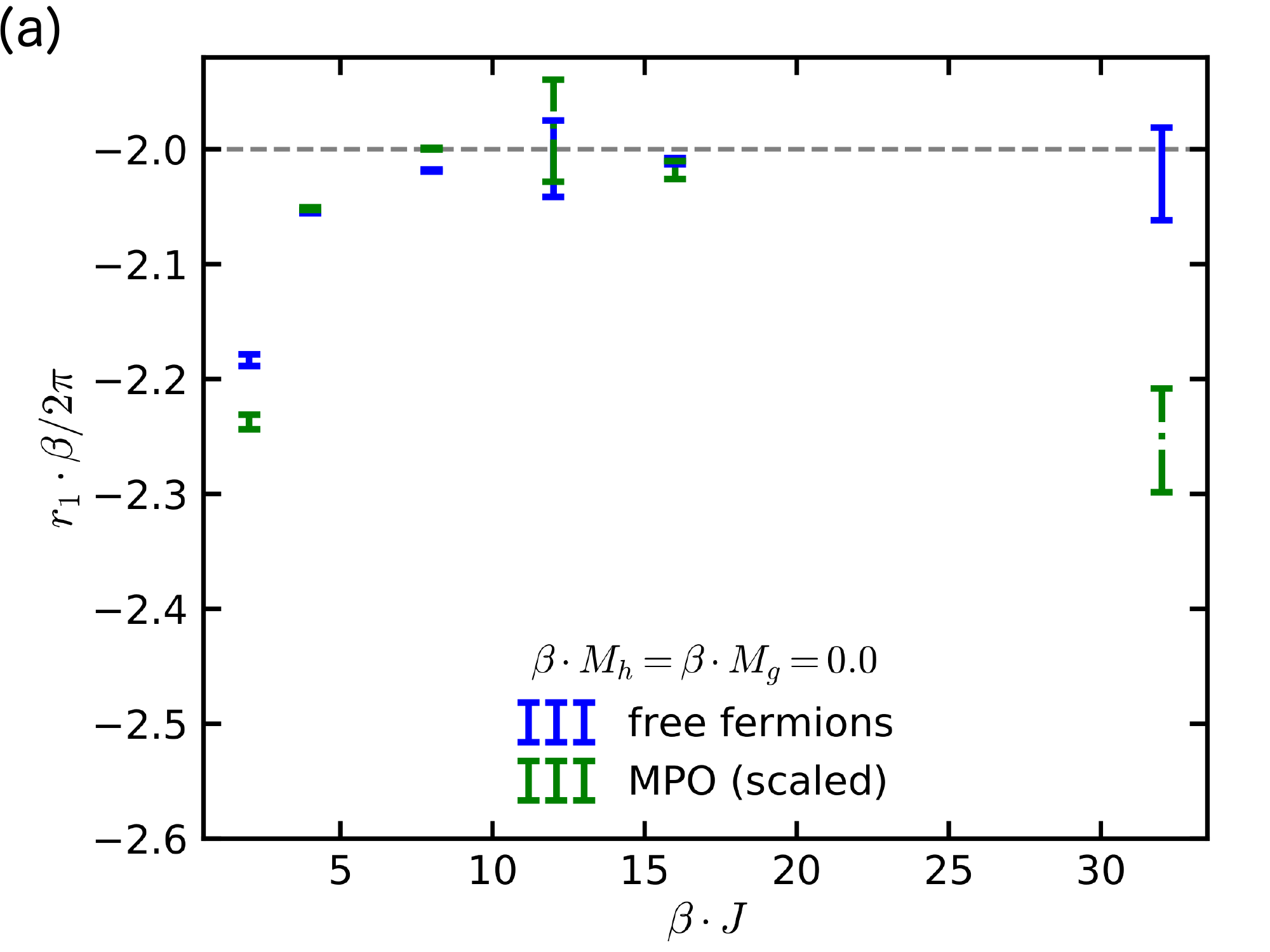}
 \includegraphics[width=0.49\textwidth]{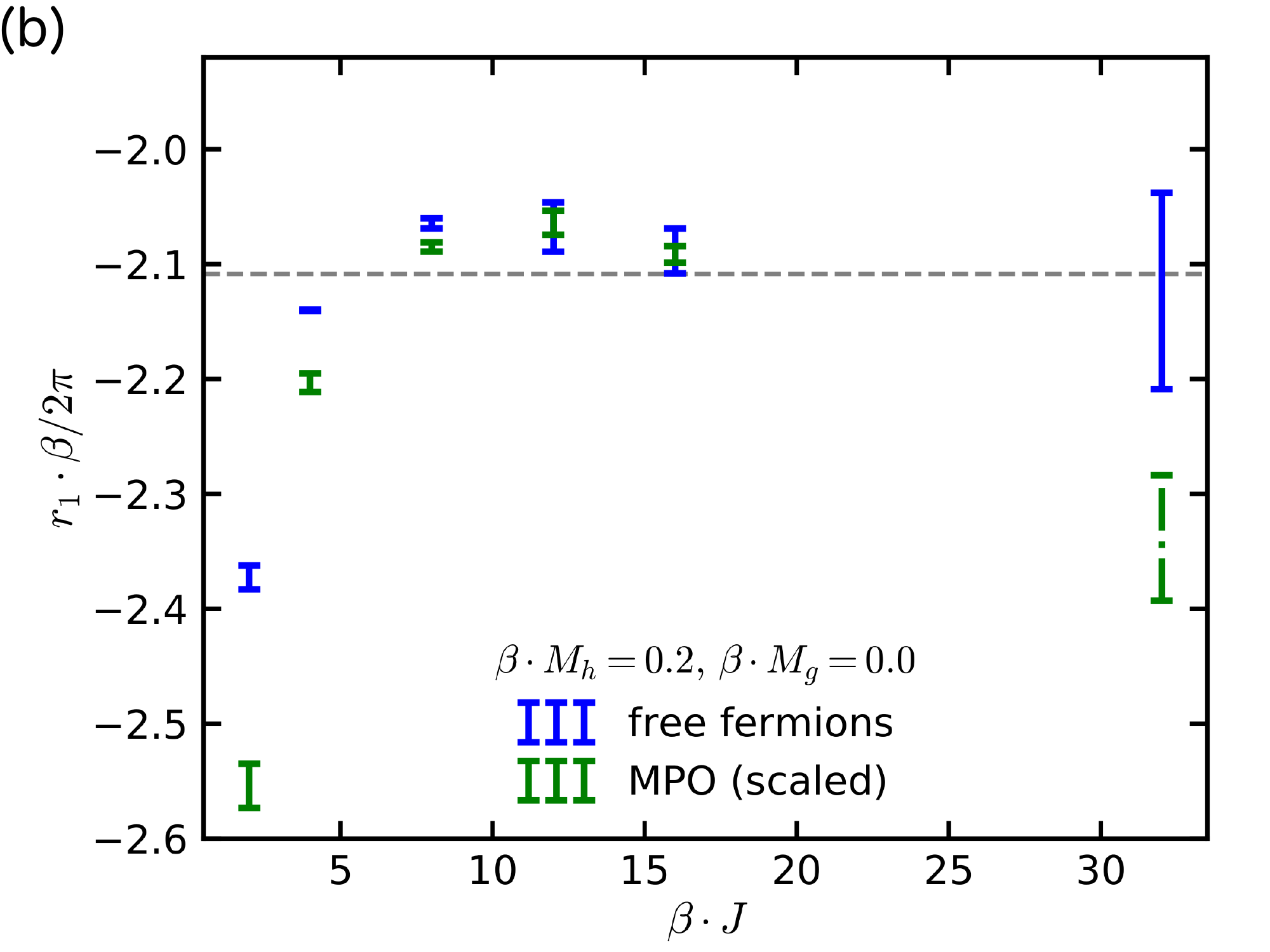} 
 \caption{Extraction of the residue of the first transient in the continuum limit based on the analytical result (blue error bars) and MPO simulation (green error bars). Panel (a) is at the critical point and panel (b) is calculated for finite transverse mass in the ferromagnetic phase. Grey lines represent the analytical result. Based on the free fermion mapping, the MPO simulations are rescaled to take the finite chain length for a proper normalization into account.}
 \label{fig:residues_int}
\end{figure}

In addition, fig.\,\ref{fig:residues_int} shows the corresponding residues of the first transient in the continuum limit. 
Panel (a) is based on the Prony analysis for the critical Ising chain, while panel (b) is for the ferromagnetic phase with the same parameters as before in this section. 
The values of the residues and their uncertainties are calculated identically to the procedure for the pole location. 
An example of this method is shown in fig.\,\ref{fig:PoleAnalyses} (b), where the coefficients of the complex exponentials in Prony are multiplied with the inverse time dependence (because of their decay in time).
The residue is then calculated as the mean value of sufficient late time windows.
From fig.\,\ref{fig:residues_int} it is visible that the residues approach the analytical value of the continuum limit (calculated from eq.\,\eqref{eq:TFIMcorrelator}) with increasing values of $\beta J$.
Although the numerical resolution is not optimal, there seems to be a clear shift in the data between the critical case ($\beta M_h=0$, (a)) and the ferromagnetic phase ($\beta M_h=0.2$, (b)).
This suggests that the Prony analysis is sensitive enough to capture the effect of the finite transverse mass, and importantly, shows that for large enough $\beta J$ the result is indeed a signal in the QFT regime.
In more detail, the data in fig.\,\ref{fig:residues_int} also shows that the residue for very large values of $\beta J$ is not correctly captured in the MPO simulation, most likely because of finite size effects and short time windows.

\subsection{The non-integrable case}

\begin{figure*}[!t]
\centering
 \includegraphics[width=0.49\textwidth]{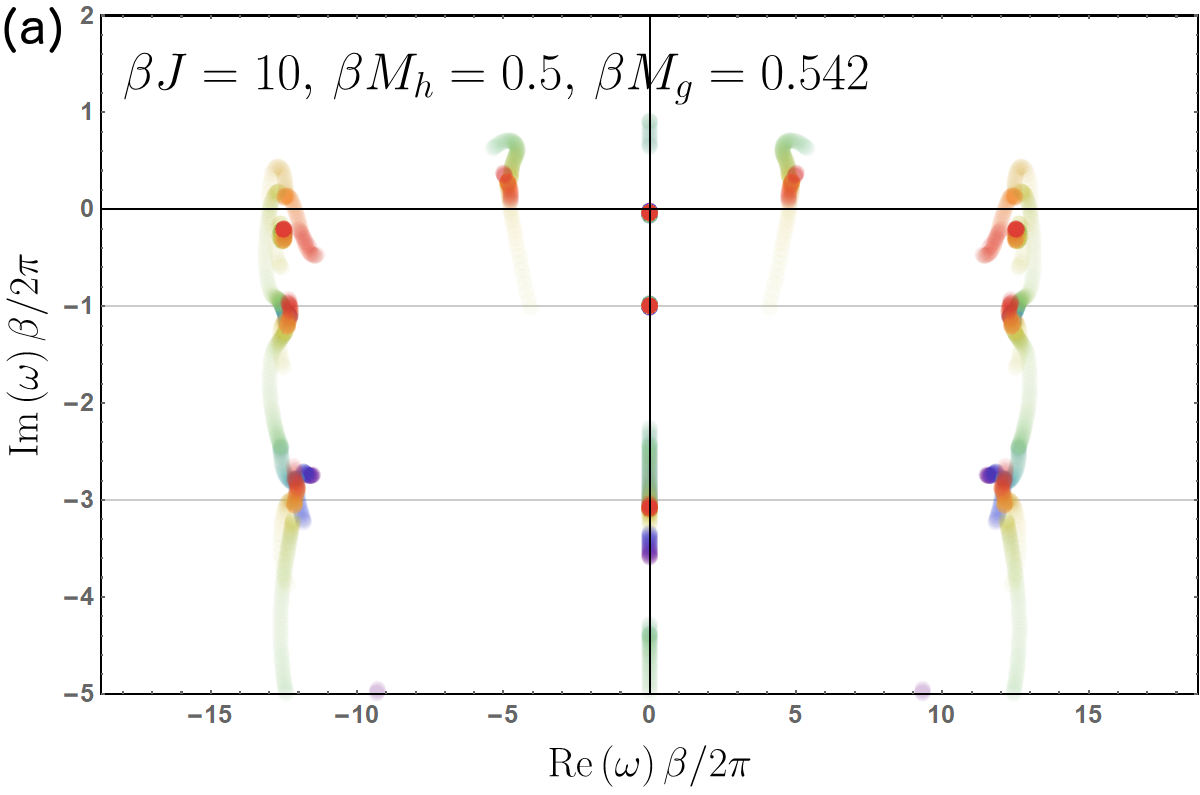} 
 \includegraphics[width=0.49\textwidth]{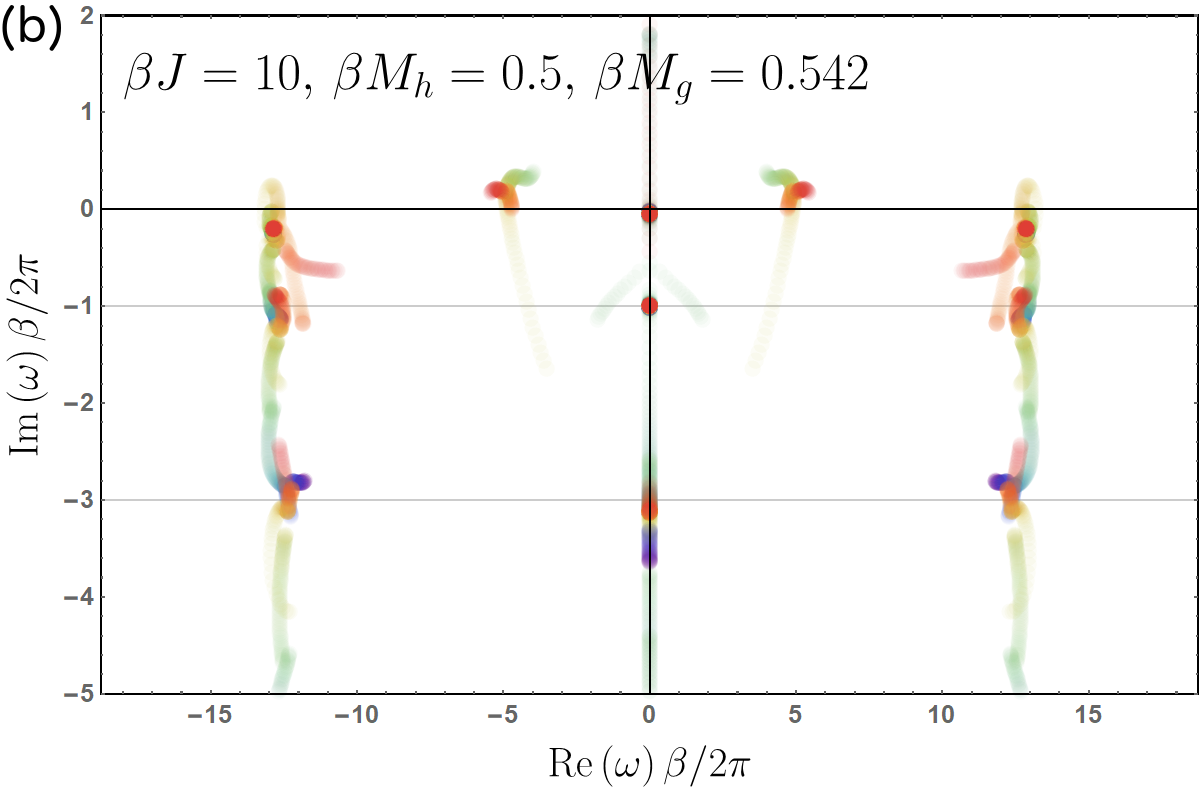} \\ 
 \includegraphics[width=0.49\textwidth]{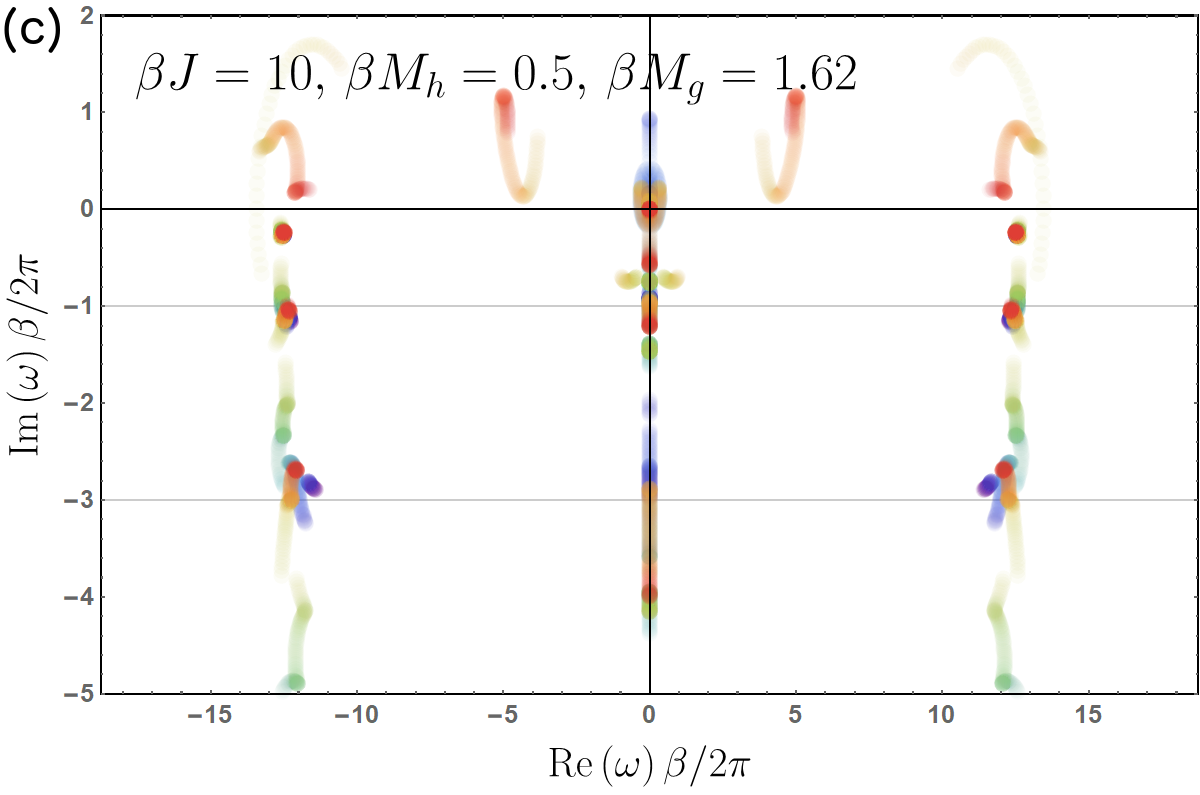}
 \includegraphics[width=0.49\textwidth]{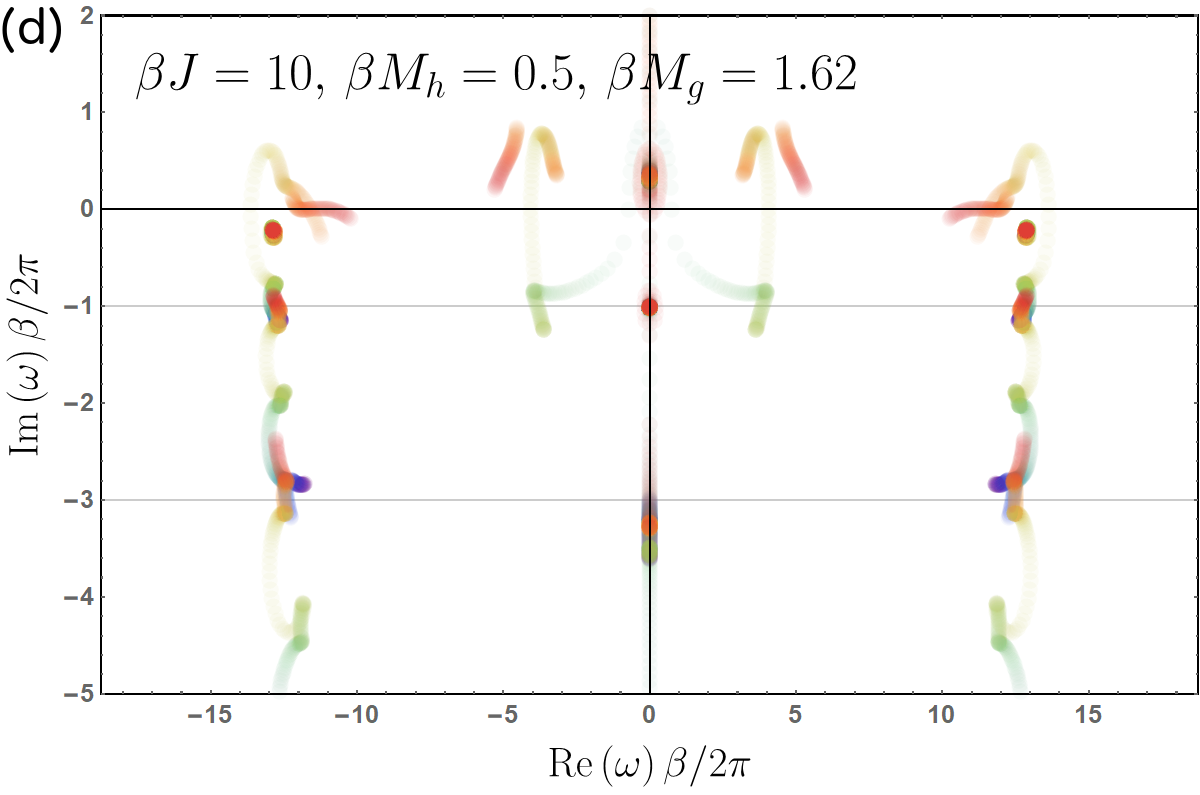} \\
  \includegraphics[width=0.4\textwidth]{time_legend_FreeFermion10.pdf} 
 \caption{Frequency analysis from Prony for the continuum limit in the non-integrable case. The left column (a,c) is in the ferromagnetic phase and the right column (b,d) in the paramagnetic phase. The imaginary axis is scaled with $\beta J/(2\pi)$. The two rows correspond to two values of the integrability breaking $\beta M_g$ at $\beta J=10$. \\
  Numerical parameters: $L=100, \chi = 200, J \delta t = 0.005,  J t_\text{max}=10$, $2^\mathrm{nd}$ order Trotter decomposition.}
 \label{fig:cont_limit_nonint}
\end{figure*}

In fig.\,\ref{fig:cont_limit_nonint}, examples of the Prony analysis of the retarded transverse correlation function are shown for two values of $\beta M_g$.
In the left column, the critical point is approached from the ferromagnetic phase $(h<1)$ and in the right column from the paramagnetic phase $(h>1)$.
Similarly to the graph in the integrable case in fig.\,\ref{fig:TFIM_correlator}, the UV branch cut is approximated by Prony through the nearly vertical line of poles at $\beta/(2\pi)\operatorname{Re}(\omega) \approx 13$.
In all examples, the first transient pole on the imaginary axis is visible at $\frac{1}{2\pi}\operatorname{Im}(\beta \omega) \approx -1$.
For the largest value of the integrability breaking $\beta M_g \approx 1.62$, the corresponding uncertainty is larger in the ferromagnetic phase (c), which is visible as a blurred set of poles in the Prony picture.
The second decaying pole is only partially visible, we therefore focus on the first pole and neglect meson frequencies for this particular study, since these would require much larger time intervals.

\begin{figure}[!h]
\centering
 \includegraphics[width=0.49\textwidth]{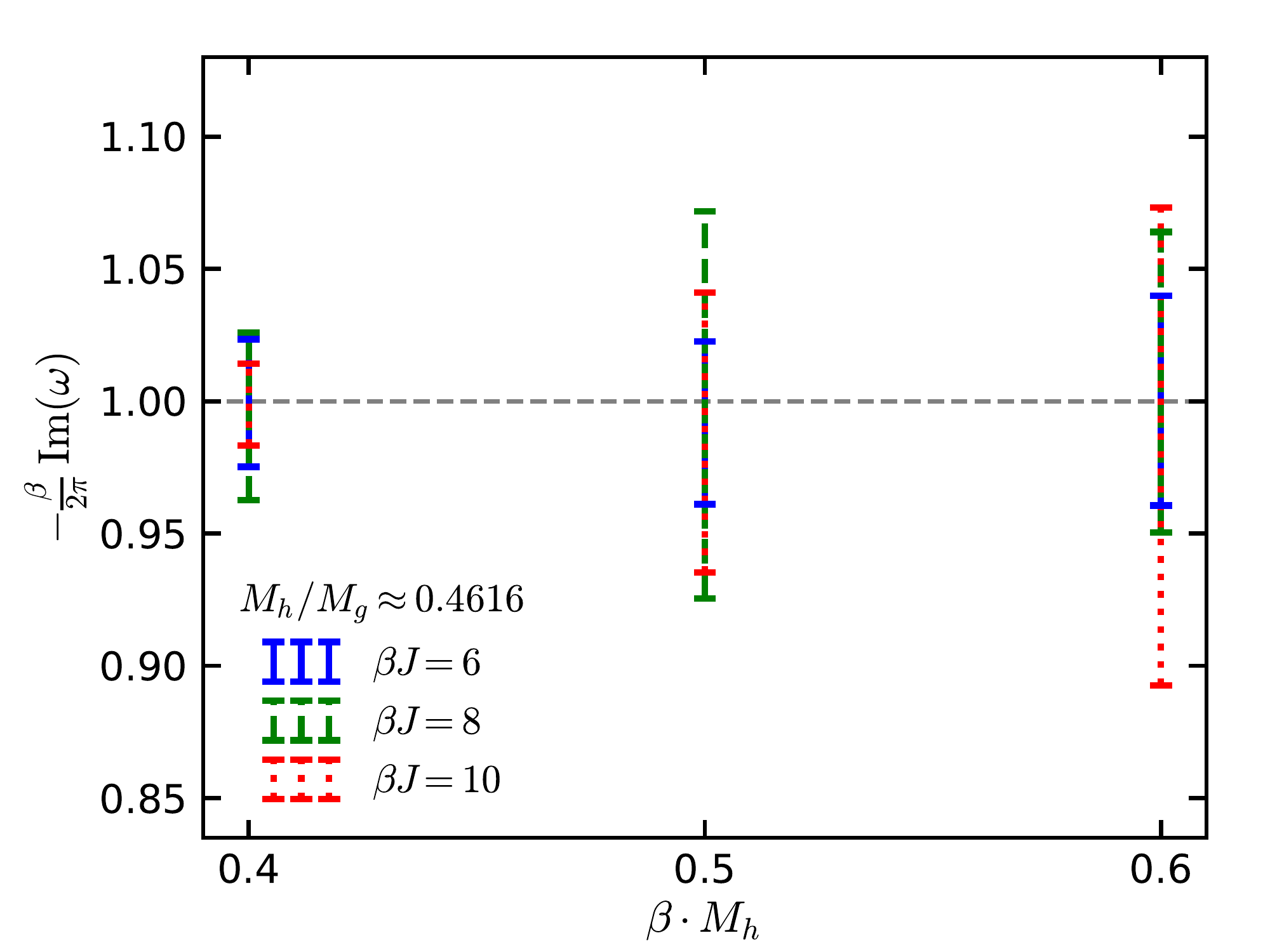}
 \caption{Extraction of the least damped transient pole in the continuum limit for the non-integrable case using TN+Prony. The critical point is approached from the ferromagnetic phase for different values of $\beta M_h$ and $\beta M_g$ such that $M_h/M_g \approx 0.4616 \equiv \text{const}$.}
 \label{fig:cont_limit_nonint_TDpoles_MhMg}
\end{figure}

The position of the extracted transient poles is shown in the main text in fig.\,\ref{fig:cont_limit_nonint_TDpoles} for the ferromagnetic (a) and paramagnetic phase (b). 
The data is generated with several parameters in the Prony method (to strengthen the robustness of our numerics).
In particular, we have chosen a cutoff value $\epsilon$ in the range $10^{-6} \le \epsilon \le 10^{-4}$ (capturing how significant modes must be to be included) and the time interval, on which the Prony analysis is applied, is within the range $75-85\%$ of the total simulation time. 
As for the integrable case described above, the Prony analysis for the shifted time windows yields an average pole position and standard deviation.
The overall uncertainty is estimated by taking the mean value of several combinations of these parameters, which then is plotted in fig.\,\ref{fig:cont_limit_nonint_TDpoles}.
When increasing the perturbation, and also when increasing $\beta J$, this uncertainty is growing up to 13\,\% in the ferromagnetic phase (a), while the maximum value does not exceed 5\% in the paramagnetic phase (b). 
This difference is related to the non-symmetric appearance of meson/particle states in the two phases, i.e.\ the continuum limit is not identical.
Fig.~\ref{fig:cont_limit_nonint_TDpoles} shows that, at fixed $\beta M_h=0.5$, the position of the thermodynamic poles is stable and consistent with the CFT prediction $\frac{1}{2\pi}\operatorname{Im}(\beta \omega) = -1$ (shown as gray dashed line), when the longitudinal perturbation is varied. 
To confirm this result, we have also checked different values of $\beta M_h$ for fixed ratio $M_h/M_g \approx 0.4616 \equiv \text{const}$,
and probed the continuum limit by taking $\beta J= \{ 6,8,10 \}$ in the ferromagnetic phase.
The results are shown in figure~\ref{fig:cont_limit_nonint_TDpoles_MhMg}.
Note that the data points at $\beta M_h = 0.5$ correspond precisely to the third set of poles in fig.\,\ref{fig:cont_limit_nonint_TDpoles}, at $\beta M_g \approx 1.08$.
For all perturbations, the resulting pole position is again consistent with the CFT result.
Within the uncertainties of the MPO simulations and numerical analysis with Prony, we therefore conclude with the prediction that the analytic structure of the retarded transverse correlation function and thermodynamic pole position does not change for (non-integrable) perturbations of the Ising CFT.

\begin{figure}[!t]
\centering
 \includegraphics[width=0.49\textwidth]{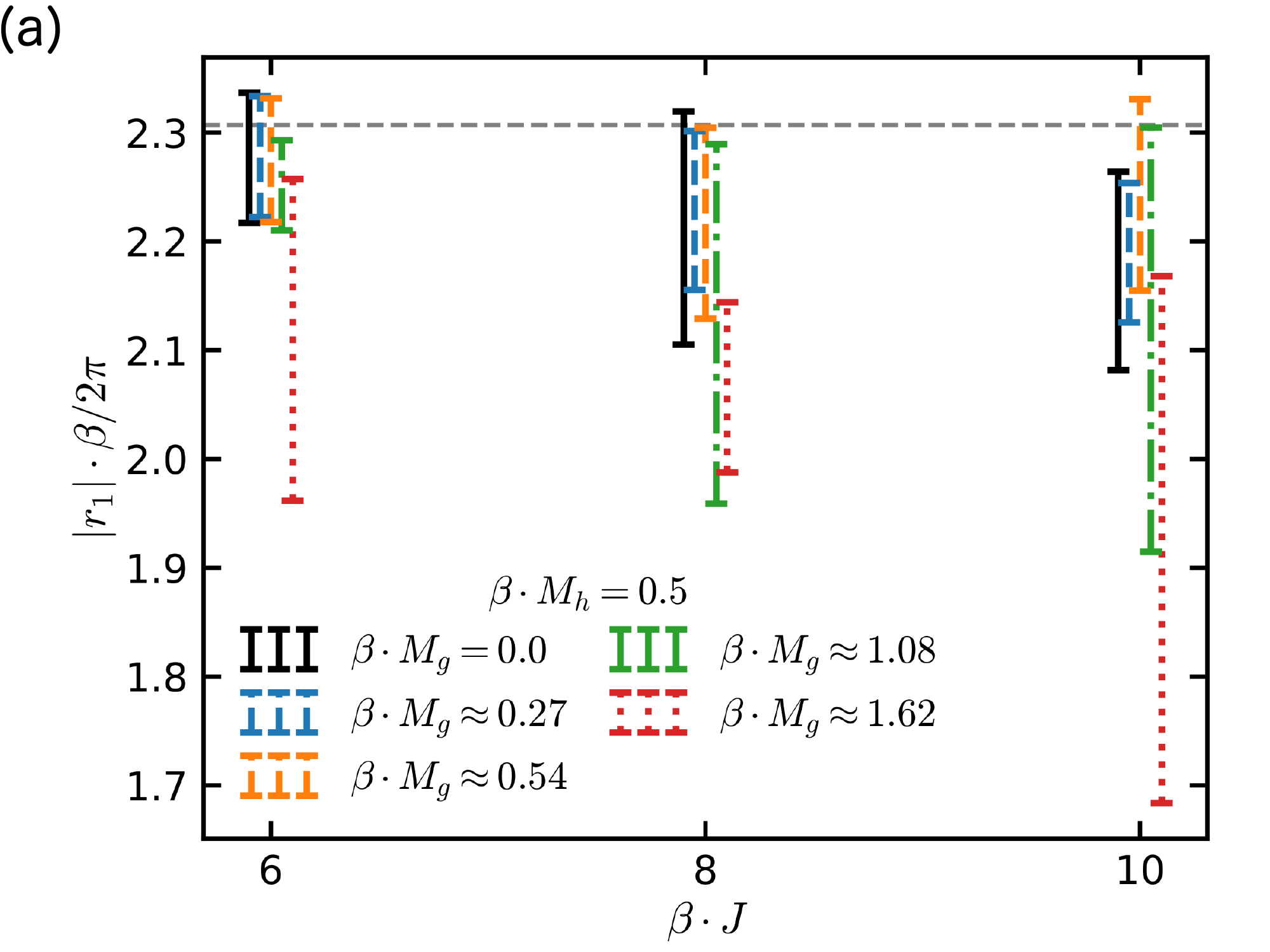}
 \includegraphics[width=0.49\textwidth]{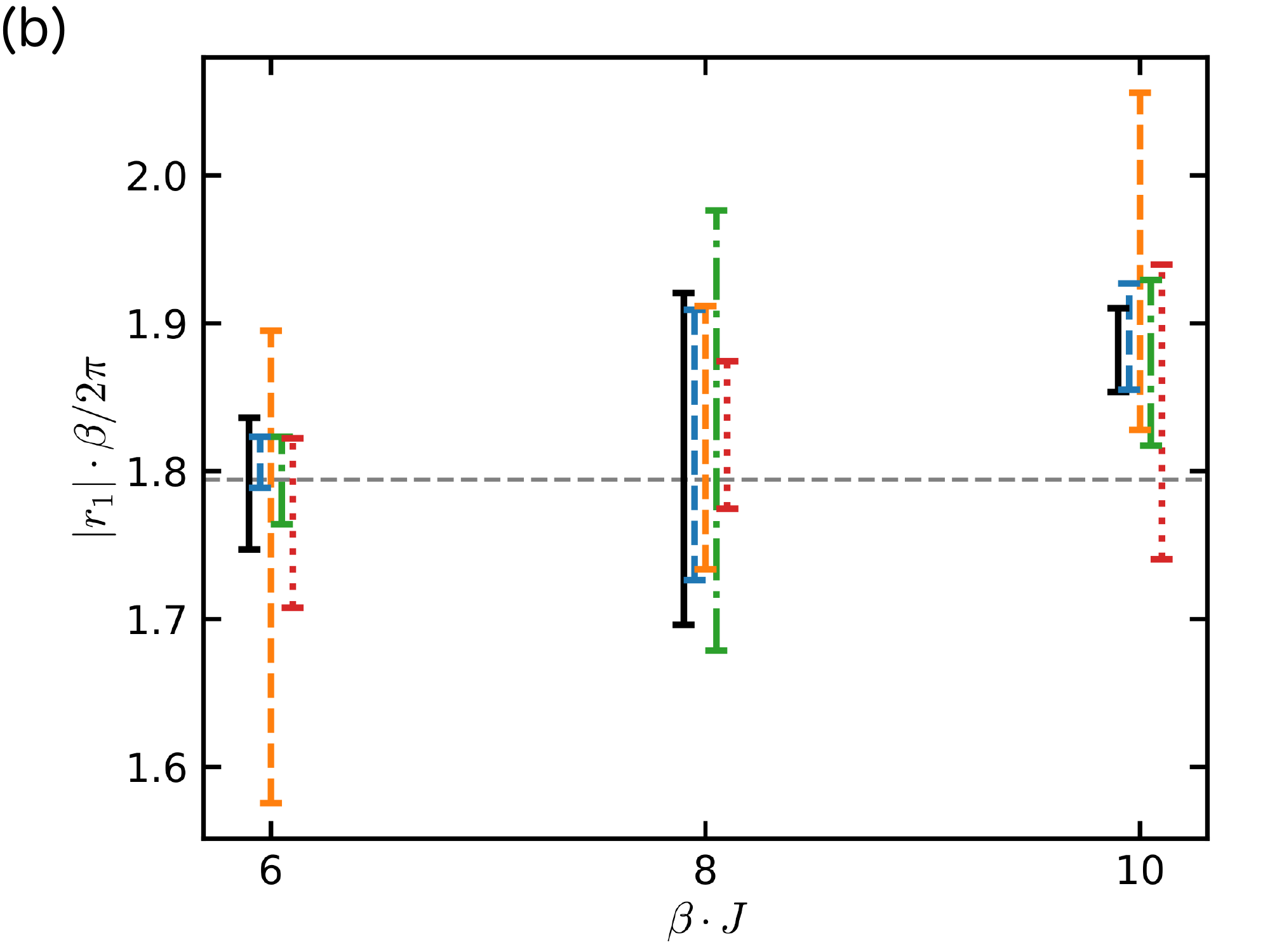} 
 \caption{Extracted residues from Prony's method for the non-integrable regime in the ferromagnetic (a) and paramagnetic phase (b) towards the continuum limit. The data are shown in dependence on $\beta J = \{6,8,10\}$ for several values of the integrability breaking $\beta M_g$. Black error bars denote the analytical value for the integrable case ($\beta M_g=0$). The curves are shown slightly displaced for graphical purposes.}
 \label{fig:residues_nonint}
\end{figure}

For the same Prony parameters and with the methodology described in the free fermion case above, we calculated the corresponding residue $r_1$ of the first decaying pole. 
Fig.\,\ref{fig:residues_nonint} shows the resulting uncertainties of $\vert r_1 \vert \beta/2\pi$ in dependence of the inverse temperature for the ferromagnetic (a) and paramagnetic phase (b).
For a comparison, the analyses was also applied to the integrable case by setting $\beta M_g=0$ (black error bars). 
Its corresponding analytical value in the continuum limit is shown as grey dashed line.
Similarly to the pole locations, the uncertainty is increasing for larger values of $\beta J$ and $\beta M_g$.
For larger longitudinal perturbations, the residues seem to exhibit the tendency to decrease in the ferromagnetic and increase in the paramagnetic phase for larger values of $\beta J$.
The orange error bars correspond to the situation when the transverse ($\beta M_h=0.5$) and longitudinal perturbation ($\beta M_g \approx 0.54$) are nearly at the same order. The corresponding data seem to be consistent with the integrable result, i.e.\ the longitudinal perturbation does not seem to influence the residue.

\section{Ground state quench\label{app:D}}

\begin{figure}[!h]
\centering
 \includegraphics[width=0.49\textwidth]{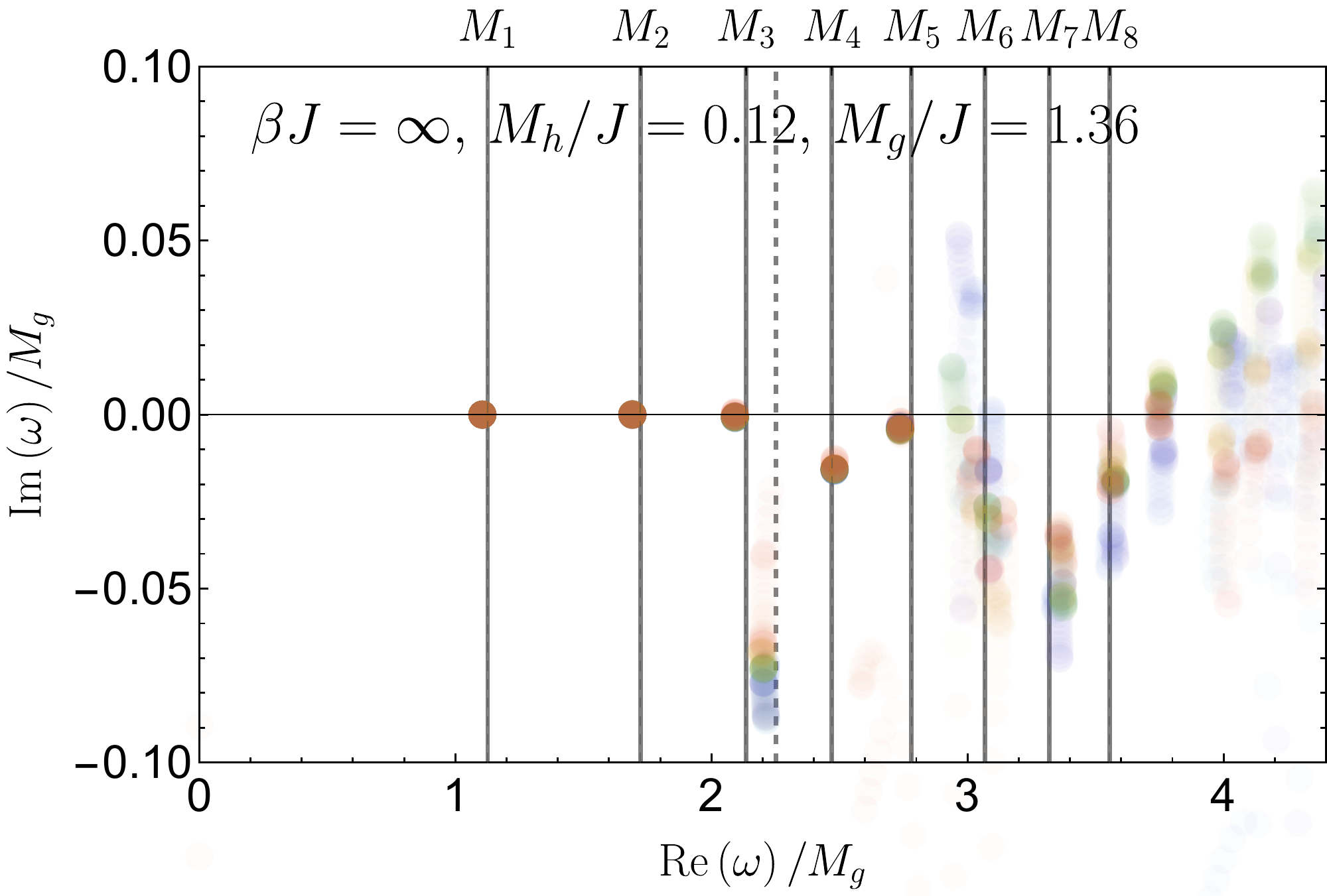}
 \caption{Prony reconstruction of the retarded correlator of $i\,\bar{\psi}\psi$ in the vacuum (i.e.\ for the ground state) in a non-integrable ferromagnetic case as in fig.\,\ref{fig:masses_nonintegrable}~(a). The solid vertical lines indicating masses are taken from ref.~\cite{Fonseca:2006au}. The dashed line indicates the continuum threshold $2 M_1$. All values are consistent with QFT.  Simulation parameters: $L=200, \chi = 170, J \, \delta t = 0.02,  J\,t_\text{max}=50$, $2^{\mathrm{nd}}$ order Trotter decomposition.}
 \label{fig:GSquench}
\end{figure}

While the present article is devoted to thermal QFT properties from spin chain simulations, we finally also want to demonstrate the applicability of our methodology to ground state quenches, i.e.\ the structure of the retarded two-point function in the vacuum.
For this purpose, we choose the set of mass parameters that were used for the meson studies in the main text (cf.\ fig.\,\ref{fig:masses_nonintegrable}~(a)).
The ground state $\ket{0}$ for this non-integrable ferromagnetic phase is calculated using the DMRG algorithm with bond dimension $\sqrt{\chi}$, such that the projector $\ket{0}\bra{0}$ can be represented as a MPO with bond dimension~$\chi$.
The MPO simulation of the correlator is then performed using the same mass parameters as for the thermal state.
The result of the Prony analysis is shown in fig.\,\ref{fig:GSquench}.

\begin{figure}[!h]
\centering
 \includegraphics[width=0.49\textwidth]{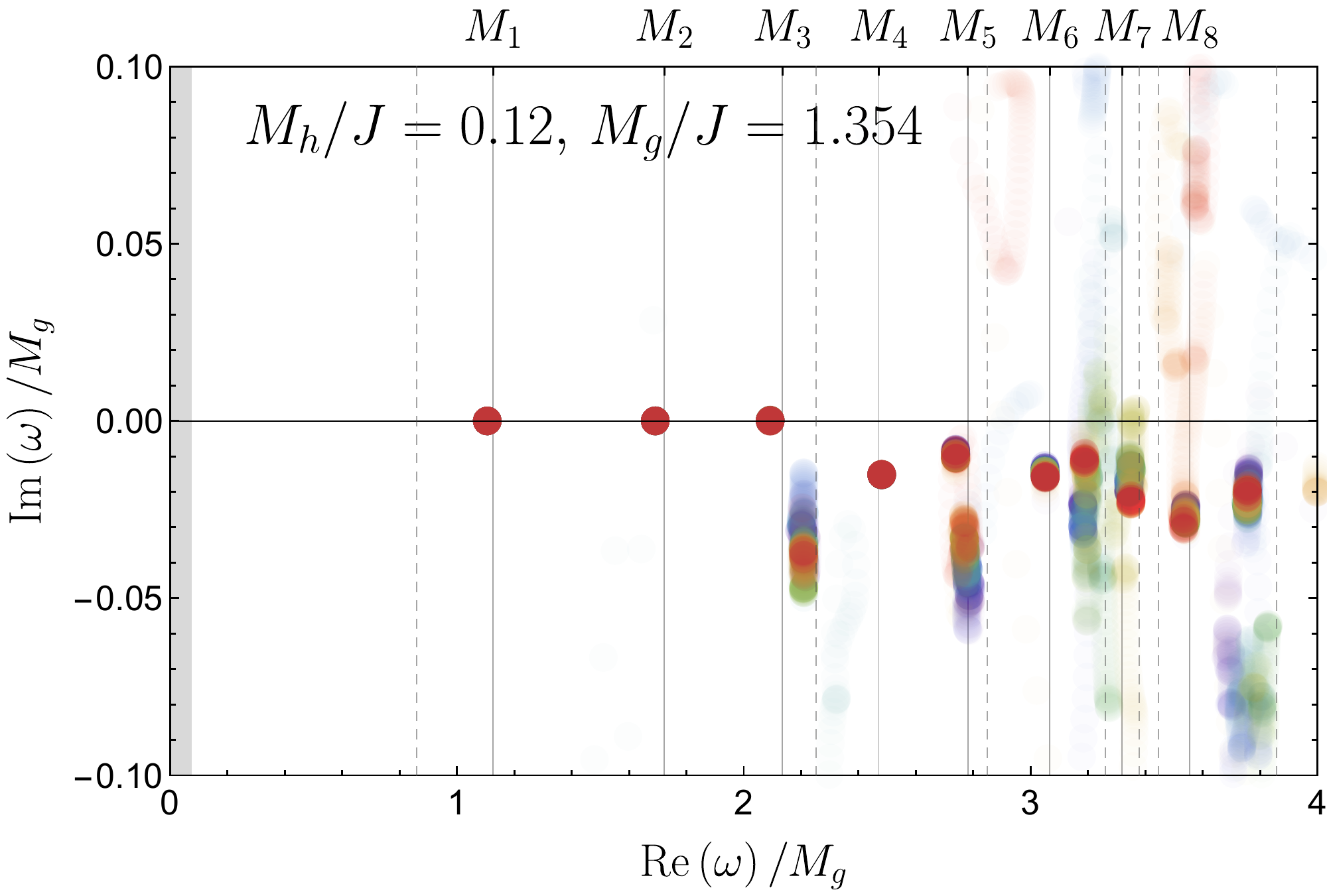}
 \caption{Prony reconstruction of the retarded correlator of $i\,\bar{\psi}\psi$ in the ground state (using MPS) in a non-integrable ferromagnetic case as in fig.\,\ref{fig:masses_nonintegrable}~(a). The solid vertical lines indicating masses are taken from ref.~\cite{Fonseca:2006au}. The dashed lines indicates different continuum thresholds. $2 M_1$ and $M_1 + M_2$ can be seen clearly, but higher ones cannot be clearly distinguished from each other and the closely lying mesons. Simulation parameters: $L=200, \chi = 150, J \, \delta t = 0.02,  J\,t_\text{max}=80, 4^{\mathrm{th}}$ order Trotter decomposition.}
 \label{fig:GSquenchMPS}
\end{figure}

Similarly to the result of the thermal state at very low temperature in fig.\,\ref{fig:masses_nonintegrable}~(a), the first three stable mesons can be identified very accurately.
While the continuum threshold $2M_1$ is visible as a branch cut stretching vertically in the Prony reconstruction, the boundary state is not seen.
The ratio of imaginary parts of the fifth and fourth meson agrees equally well with the prediction in ref.~\cite{Delfino2005Jul} as in the small temperature simulation.
The Prony reconstruction at high frequencies becomes more fuzzy, although some structures at the known masses are identifiable.
These results clearly demonstrate that the effects at higher temperature, that we discussed in the main text, indeed have a thermal origin. 
Furthermore, our findings demonstrate that results obtained from the vacuum density matrix constructed from TEBD as $\lim_{\beta \rightarrow \infty} e^{-\beta H}$ and from DMRG as $\ket{0}\bra{0}$ agree.
Instead of evolving the MPO $\ket{0}\bra{0}$, one could also evolve just $\ket{0}$ as a matrix product state (MPS). This saves us from squaring the bond dimension, enabling us to evolve the correlator for longer time and hence more accurately. The result is shown in fig.\,\ref{fig:GSquenchMPS}. The main gain is in enabling us to see the start of the $M_1+M_2$ continuum, as well as making the higher mesons sharper.

\end{document}